# RFID-Based Non-Biometric Classroom Attendance System: Proxy Attendance Detection via Weight Sensor Integration

**Furkan EGE[1], Muhsin ÖZDEMİR[1]**

[1] *Department of Management Information Systems, Institute of Social Sciences, Aydın Adnan Menderes University, Aydın, Türkiye*

## ABSTRACT

Attendance tracking in educational institutions, when conducted through traditional methods, leads to structural problems that consume instruction time and threaten academic integrity. Attendance durations spanning several minutes in primary and secondary education and exceeding ten minutes in higher education, combined with the proxy attendance problem of signing on behalf of someone else, demonstrate the need for electronic systems. Most existing electronic solutions rely on biometric authentication, which raises legal and ethical risks under the European General Data Protection Regulation (GDPR), the Turkish Personal Data Protection Law (KVKK), and the United States Family Educational Rights and Privacy Act (FERPA). Systems using RFID alone provide no built-in safeguard against proxy attendance through card transfer.

This study proposes a biometric-free IoT attendance system addressing both deficiencies. The prototype consists of an RFID module, RFID cards, weight sensors, a Bluetooth module, and an Arduino UNO microcontroller. After the student presents their RFID card, the weight sensor measurement is compared against a statistical reference range of 350 individuals (aged 18–22) compiled from three Kaggle datasets; no personal biometric data is recorded. A Python-based GUI performs student management, course tracking, and CSV-based reporting via Bluetooth.

Qualitative tests in conditions close to a real classroom have shown that the RFID reading, weight verification, Bluetooth communication, and GUI modules operate in an integrated manner as expected. The proposed system offers a low-cost and reproducible solution that aims to reduce proxy attendance without storing biometric data.

# 1. INTRODUCTION

Education is the process of deliberately bringing about desired changes in an individual's behavior through their own experience, and educational institutions are the organizations where this process takes place and where students are equipped with knowledge, skills, and attitudes (Ertürk, 1997: 12). Attendance tracking, an indispensable part of this process, is directly related to students' academic success and is also a critical component that supports the sustainable operation of educational institutions from an administrative perspective. However, the traditional signature-based or verbal attendance methods widely used for recording student attendance exhibit a structure that consumes valuable instruction time and carries serious reliability issues.

The verbal attendance method currently applied in primary and secondary education institutions is carried out by the teacher calling out student names one by one and students responding. This process takes 3 to 6 minutes depending on class size. In higher education institutions, the signed attendance sheet method is widely used; due to the high number of students, this duration can exceed 10 minutes (Sezdi and Tüysüz, 2018). In addition to this time loss, the signed attendance method also brings along the proxy attendance problem manifested as "signing on behalf of someone else."

Numerous electronic attendance systems have been proposed to address these problems. Based on different technologies such as fingerprint recognition, face recognition, iris scanning, voice biometrics, NFC, Bluetooth, Wi-Fi, and QR codes, these systems aim to automate the attendance process, saving time while preventing proxy attendance (Ishaq and Bibi, 2023). However, the vast majority of these approaches require the collection of biometric data. Such data can cause permanent damage in the event of a data breach (Kindt, 2013) and is regulated as special category personal data or as a protected education record under GDPR (European Union, 2016), the Turkish KVKK, and the U.S. FERPA (U.S. Department of Education, 1974); this situation significantly restricts the applicability of such systems in educational institutions.

On the other hand, systems based solely on Radio Frequency Identification (RFID) technology cannot provide sufficient verification assurance on their own (Duroc, 2022). A student entrusting their RFID card to another person constitutes the most fundamental weakness of these systems.

This study proposes an integrated attendance system that simultaneously addresses two fundamental deficiencies: the requirement for biometric data and insufficient assurance against proxy attendance. The developed system brings together RFID technology, weight sensors, a Bluetooth module, and an Arduino UNO microcontroller. The instantaneous measurement obtained from the weight sensor is compared against intervals derived from statistical datasets of 350 individuals aged 18–22; since this approach does not require biometric data storage, it significantly reduces the data protection obligations in the context of GDPR, KVKK, and FERPA.

The main contributions of this paper can be summarized as follows: (i) a privacy-focused attendance architecture that does not require biometric data; (ii) a novel proxy attendance detection mechanism that integrates RFID identity verification with weight-sensor-based physical presence verification; (iii) a low-cost and reproducible IoT prototype tested in an environment close to real classroom conditions; and (iv) a systematic data analysis of the weight distribution for the 18–22 age group.

The remainder of the paper is organized as follows. Section 2 reviews the literature and identifies the research gap. Section 3 introduces the system's design decisions, component selections, and architectural structure. Section 4 presents the statistical basis of the weight verification thresholds. Section 5 details the physical assembly of the prototype and the software development process, while Section 6 reports the results of module and integrated system tests. Section 7 discusses the proposed system, and Section 8 presents conclusions and future research directions.

## 2. RELATED WORK

Research on electronic attendance systems can be examined under three main categories according to the authentication technology used: biometric-based systems, non-biometric systems, and hybrid approaches. This section reviews representative studies in these categories and identifies the existing research gap.

## 2.1. Biometric-Based Systems

Biometric authentication constitutes the broadest category of electronic attendance studies in the academic literature. Nawaz et al. (2009) developed one of the first comprehensive systems to electronically record student attendance by placing fingerprint sensors at classroom entrances. Koçak and Yeleç (2017) presented a system based on a biometric fingerprint reader that achieved high user satisfaction in usability tests. Soewito et al. (2016) proposed a smartphone-based system that combines fingerprint verification with voice recognition, achieving a 95% match rate in fingerprint verification. Kamelia et al. (2018), by integrating a fingerprint module with GPS, performed real-time location and attendance tracking; verification was completed in an average of 1.39 seconds.

Studies based on face recognition technology also represent a substantial strand of research in this field. Sawhney et al. (2019) developed a real-time face-recognition-based automatic attendance system using PCA and CNN algorithms. Lukas et al. (2016) extracted facial features using DWT and DCT methods and obtained an 82% success rate in 148 recognition operations. Turan and Doğan (2024) developed a recent face-recognition-based student tracking system. Dey et al. (2014) established a voice biometrics platform using i-vector-based speaker recognition modeling and reported a 94.2% accuracy rate in a two-month test. Okokpujie et al. (2017) presented an iris scanning system integrated with a web interface.

Walia and Jain (2016), through a comprehensive review of fingerprint-based biometric attendance systems, drew attention to the fundamental limitation of such systems: the storage of immutable personal characteristics in central databases. Regulations such as GDPR, KVKK, and FERPA regulate such data as special category personal data or as protected education records; therefore, deploying such systems in educational institutions creates serious legal obligations (Kindt, 2013).

## 2.2. Non-Biometric Systems

Non-biometric systems rely on optical, wireless signal, and RFID technologies for authentication. Among studies based on optical and near-field communication, Masalha and Hirzallah (2014) proposed a QR code system in which attendance is completed via smartphone scanning and multi-factor authentication. Baykara et al. (2017) developed a mobile attendance system based on NFC tags and using Google Maps for location verification. Benyó et al. (2012) presented a hybrid solution that integrates NFC cards with fingerprint verification.

Wireless signal-based approaches also represent a substantial strand of work within this category. Puckdeevongs et al. (2020) established a Bluetooth-based attendance system performing in-class positioning based on RSSI values obtained from BLE beacons. Bhalla et al. (2013) proposed a system based on a teacher's mobile phone using Bluetooth MAC address collection. Banepali et al. (2019) established a Wi-Fi-based system achieving over 94% accuracy over WLAN. Küçük et al. (2018) presented a school bus tracking system using BLE and Firebase.

Passive RFID technology occupies an important place in the electronic attendance literature due to its low cost and its not requiring an external power source (Duroc, 2022). Chiagozie and Nwaji (2012) developed one of the first comprehensive RFID-based systems. Sezdi and Tüysüz (2018) presented a web-based RFID attendance system capable of integrating with institutional information systems. Aydın and Dalkılıç (2018) revealed the relationship between student attendance rates and academic achievement through data mining. Pala (2008) and Özcan et al. (2018) developed, respectively, an RFID-based e-attendance system and an integrated RFID+BLE system. In all such systems that use only RFID, the proxy attendance problem persists when the card is transferred.

## 2.3. Hybrid Systems and Identification of the Research Gap

To address the proxy attendance problem, some studies have combined RFID with biometric verification. Uludağ and Uçar (2018) established a smart classroom system that integrates RFID with fingerprint verification on a Raspberry Pi-based IoT infrastructure. Akbar et al. (2018) proposed a system combining RFID with OpenCV-based face recognition. Although these hybrid approaches largely solve the proxy attendance problem, they do not eliminate the obligation to store biometric data.

The literature review reveals that existing studies exhibit at least one of two fundamental deficiencies: they either require the collection of biometric data and bring with them the legal risks under GDPR, KVKK, and FERPA, or they use RFID alone and fail to provide assurance against the proxy attendance problem. This study aims to fill this gap with a novel IoT architecture that integrates RFID identity verification with weight-sensor-based physical presence verification without collecting biometric data.

## 3. SYSTEM ARCHITECTURE AND DESIGN

This section presents the design decisions, component selections, and architectural structure of the developed system. Unlike the physical assembly and software development process in Section 5, this section addresses the theoretical framework of the system and the operational logic of the novel proxy attendance detection mechanism.

### 3.1. General System Architecture

The proposed system is built on an integrated IoT architecture consisting of three complementary layers: the sensing layer, the processing layer, and the management layer. In the sensing layer, the passive RFID module identifies the student while the weight sensors placed under the chair independently verify physical presence. In the processing layer, the Arduino UNO microcontroller processes data from both sources in real time to generate the attendance decision; in the management layer, data transmitted to the computer via the HC-06 Bluetooth module is managed through the Python GUI. Figure 1 shows the Fritzing circuit diagram showing the connections of all components.

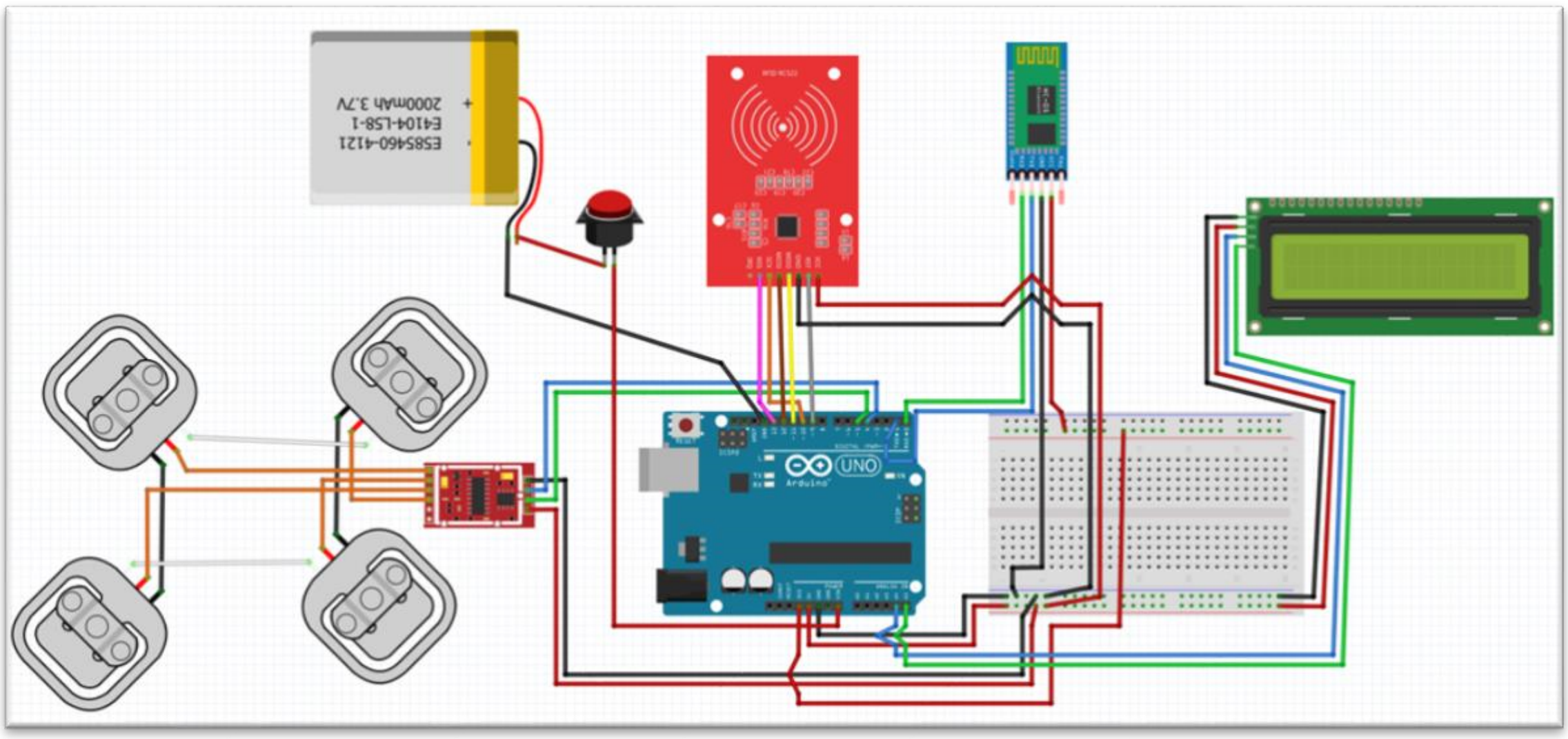

*Figure 1. Fritzing circuit diagram of the system (Arduino UNO, RC522, HX711, HC-06, LCD).*

This architecture has been deliberately designed to not require a central server or cloud connection. This choice provides two important advantages: the system continues to operate in environments where the internet connection is intermittent, and data privacy is structurally safeguarded by ensuring that student data does not leave the institution's physical boundaries.

## 3.2. Hardware Components

### *3.2.1. Arduino UNO Microcontroller*

The processing core of the system is the ATmega328P-based Arduino UNO board (Badamasi, 2014; Köhli et al., 2024). This board, which has 14 digital input/output pins, 6 analog input pins, and 32 KB of program memory (Atmel Corporation, 2015), enables the simultaneous management of the RFID module, weight sensor amplifier, Bluetooth module, and LCD display.

### *3.2.2. RC522 RFID Module and MIFARE Cards*

For student identity verification, the RC522 RFID module operating at 13.56 MHz was selected (NXP, 2016). This module is used with MIFARE Classic 1K cards compliant with the ISO/IEC 14443 standard (ISO, 2018; NXP, 2018). The RC522 communicates with the Arduino via the SPI protocol and has a typical reading distance of approximately 3–5 cm. The use of passive RFID technology enables the tags to operate without requiring external power (Want, 2006).

### *3.2.3. Load Cells with HX711 ADC*

Physical presence verification is carried out with four half-bridge load cells of 50 kg capacity placed under the chair. Figure 2 presents a view of the load cells.

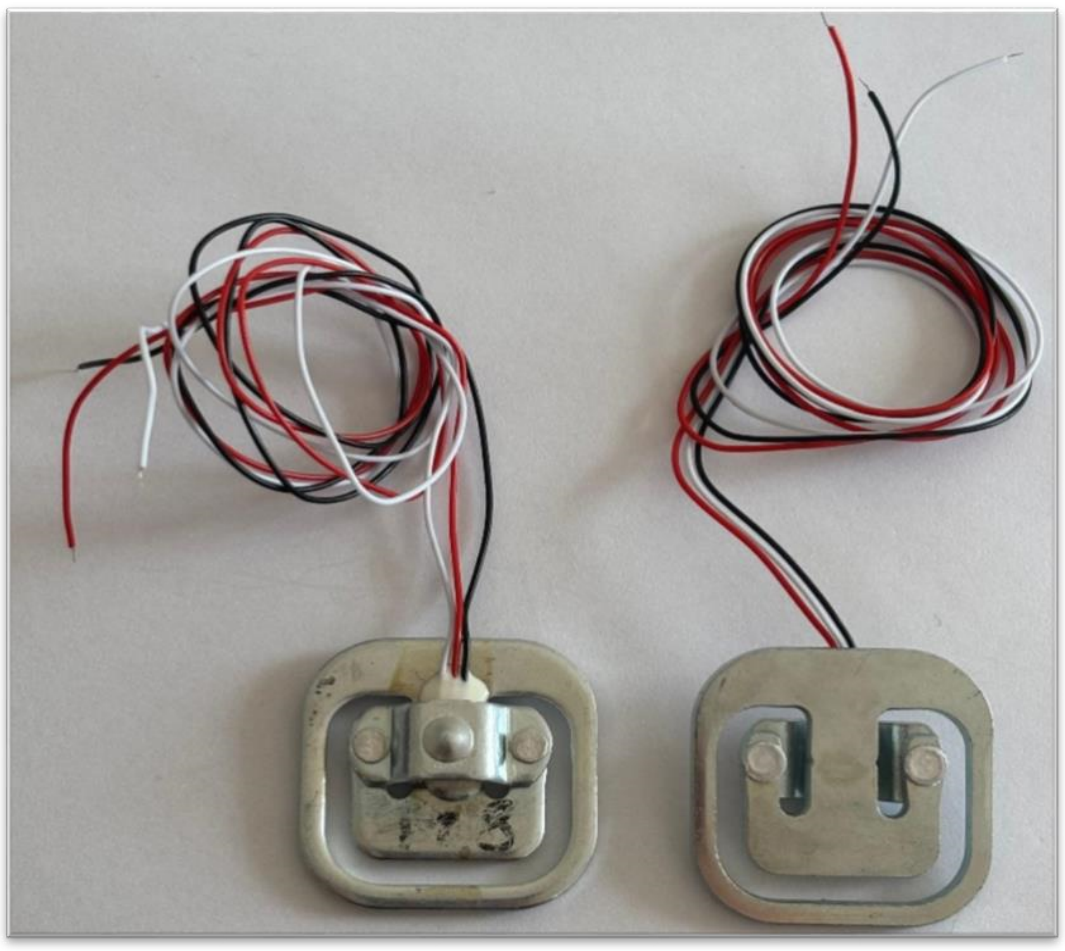

*Figure 2. Half-bridge load cell with a capacity of 50 kg.*

The HX711 24-bit ADC module converts the analog signal into digital data and transmits it to the Arduino (SparkFun, September 2016; AVIA Semiconductor). Connecting the four sensors in a Wheatstone bridge configuration increases measurement stability.

### *3.2.4. HC-06 Bluetooth Module*

Data transmission between the Arduino and the computer is carried out wirelessly through the HC-06 serial Bluetooth module over the UART protocol. In the prototype, this module enables the transfer of data received from the Arduino to the Python-based graphical user interface, providing a short-range and low-cost communication infrastructure between the system's processing and management layers. The HC-06 module operates in the 2.4 GHz frequency band and can provide effective communication at a range of approximately 10 meters in open environments (Research Design Lab, n.d.). Figure 3 shows a view of the HC-06 Bluetooth module.

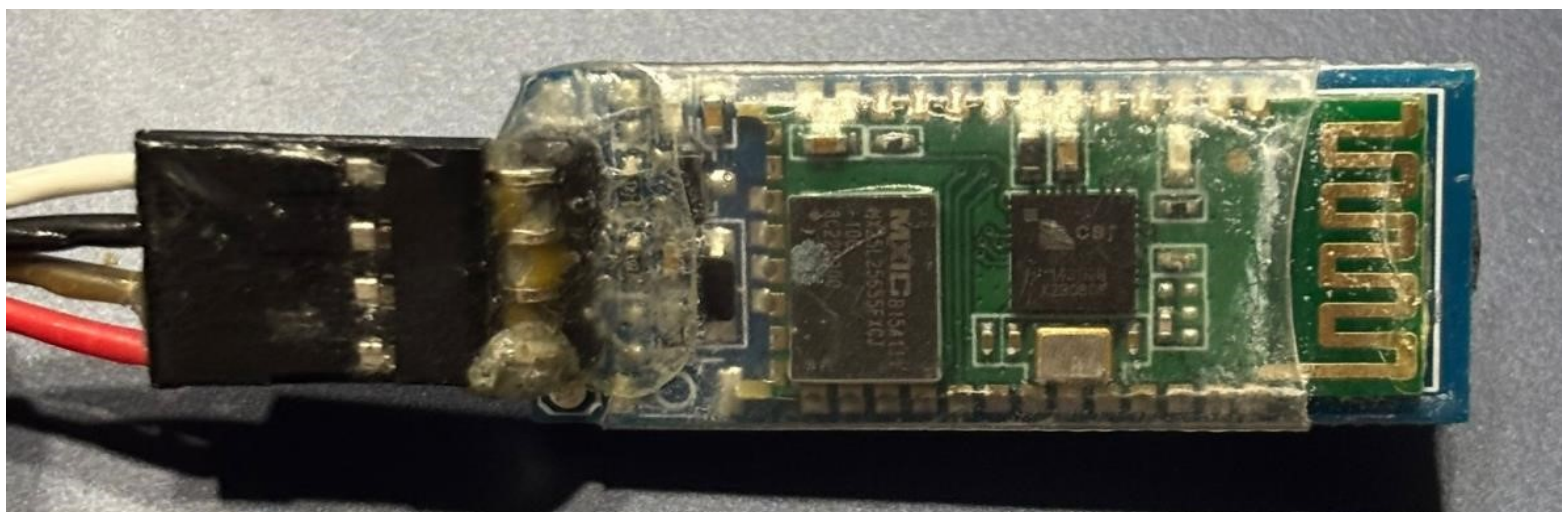

*Figure 3. HC-06 Bluetooth module.*

### *3.2.5. 16×2 I2C LCD Display*

A 16×2 LCD display with I2C interface was used to provide instant feedback to students. Since the I2C protocol requires only two communication lines, the Arduino's pin usage is significantly reduced (HandsOn Tech., 2018). Figure 4 presents a view of the LCD display. The screen instantly visually reports conditions such as successful scans, unrecognized cards, incorrect time, and attendance confirmation.

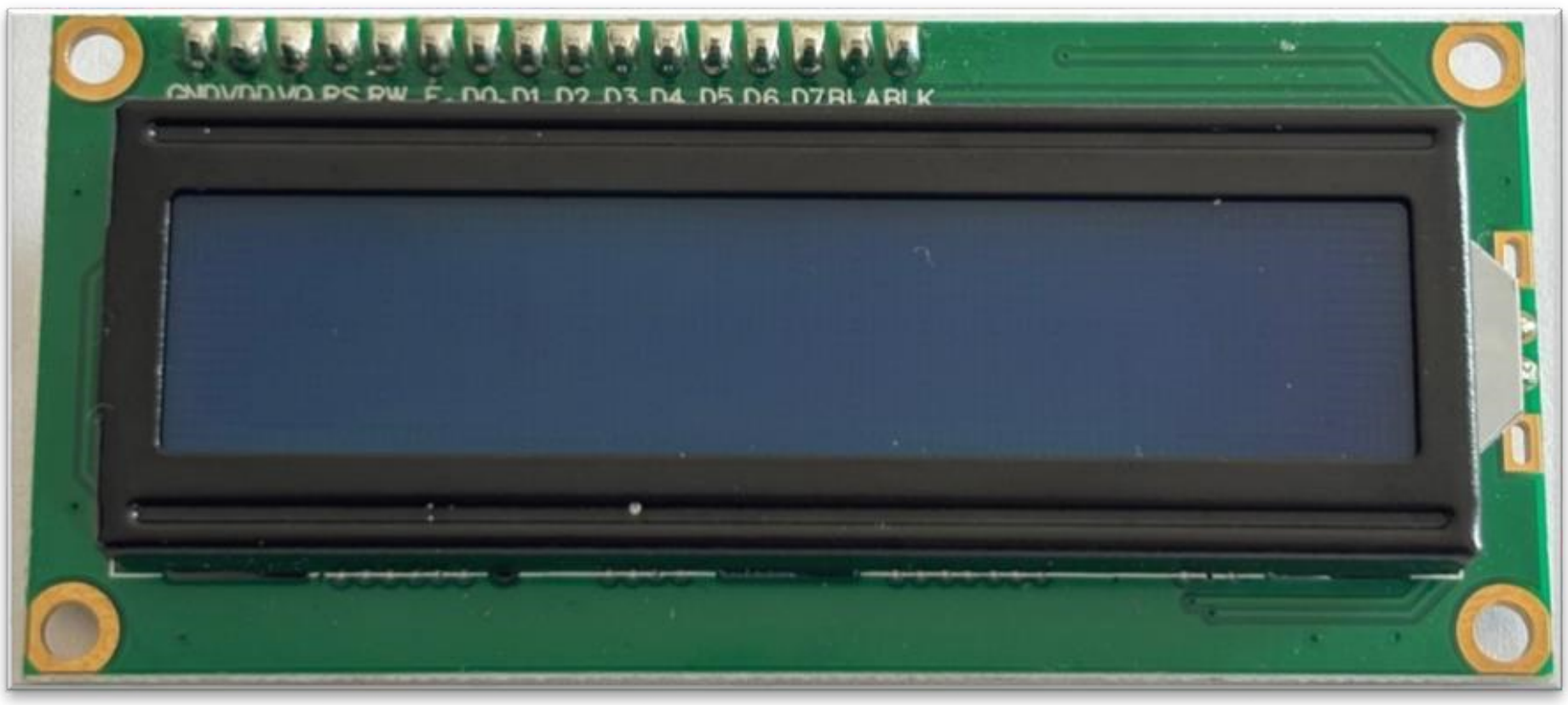

*Figure 4. 16×2 I2C LCD display.*

## 3.3. Software Architecture

### *3.3.1. Arduino Embedded Software*

The embedded software running on the Arduino board was developed in the Arduino IDE environment (Arduino, 2024). The software loop sequentially performs three fundamental tasks: obtaining the card ID from the RFID reader, processing the instantaneous reading value from the weight sensor, and combining these two pieces of data to generate the attendance decision. A retry logic for card reading and a tolerance-based filtering algorithm for sensor noise have been implemented.

### *3.3.2. Python GUI Application*

The management interface on the computer side was developed in Python in the PyCharm IDE environment. The application performs the functions of student record management, course and time definition, processing attendance data over Bluetooth, and storing records in CSV format. CSV-based storage was deliberately chosen to facilitate integration with institutional information management systems.

## 3.4. Proxy Attendance Detection Mechanism

The proxy attendance detection mechanism, which constitutes the original contribution of the system, is based on a two-stage logic that cross-compares RFID-based identity verification with weight-sensor-based physical presence verification.

In the first stage, the system matches the UID obtained from the RFID card against the records on the computer, verifying whether the card is registered, whether the student is enrolled in the relevant course, and whether the scanning time coincides with the course's time slot. When the match is successful, the message "Welcome, [Full Name]" is displayed on the LCD screen; when verification fails, a "Student Not Found" warning is issued. Figure 5 shows the flowchart of this decision logic.

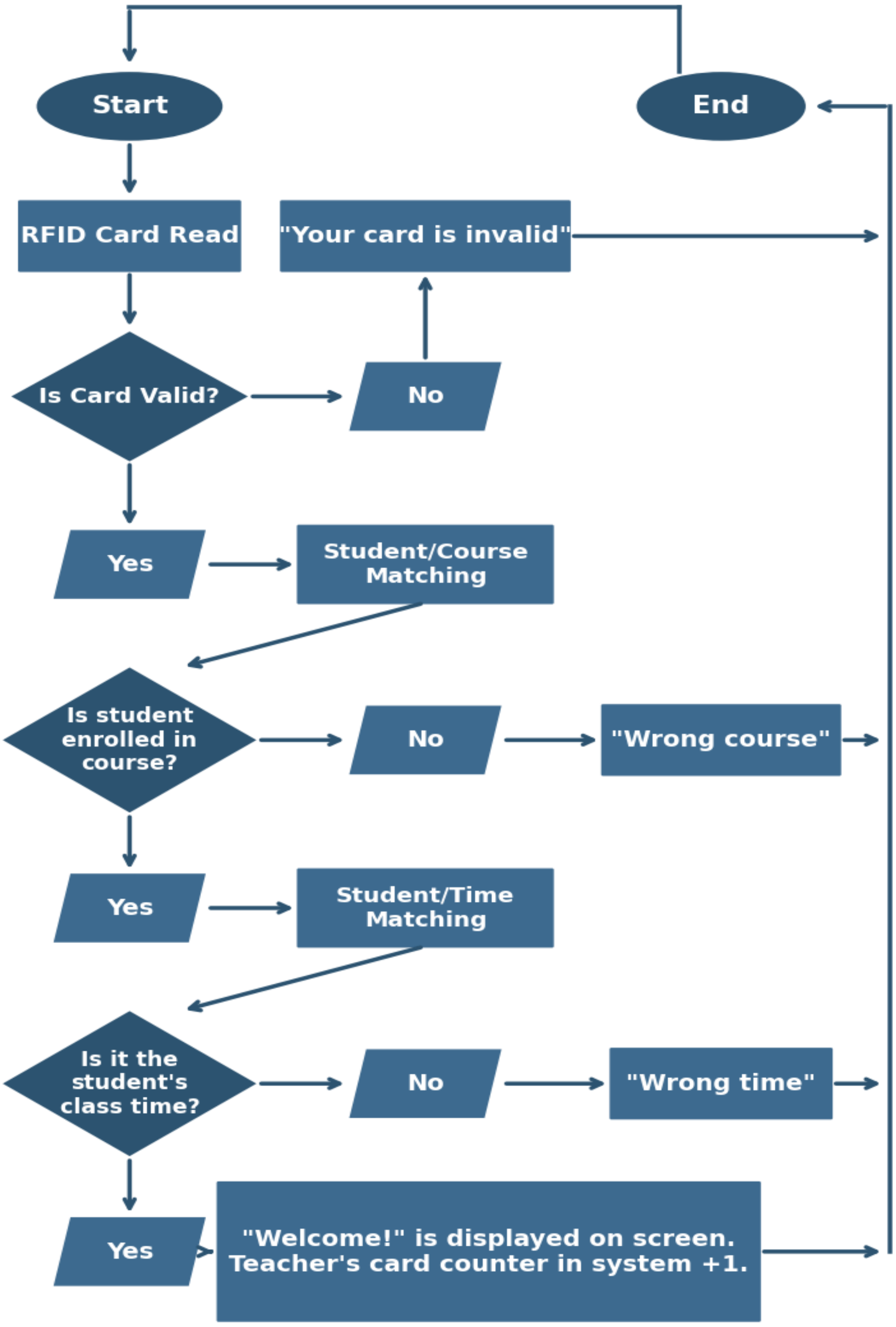

*Figure 5. Flowchart of RFID card scanning and verification.*

In the second stage, the weight value obtained from the corresponding chair is compared against gender- and age-based ranges derived from the statistical dataset. Weight values are not recorded in the system; they are used only for one-time comparison.

Because this approach does not require the storage of individual biometric features, it significantly reduces the data protection obligations in the context of GDPR, KVKK, and FERPA. Figure 6 provides the flowchart of the weight sensor verification process.

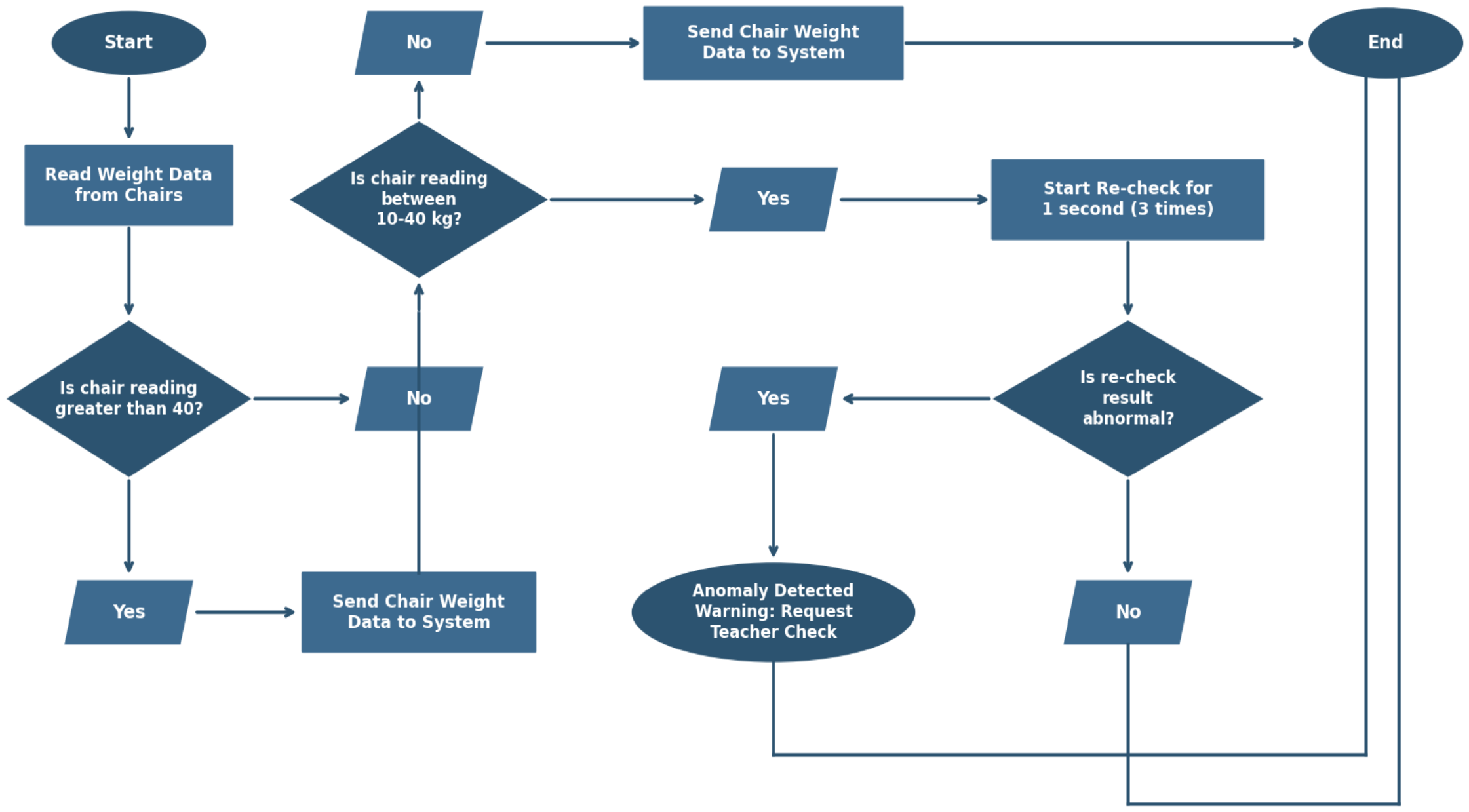

*Figure 6. Flowchart of weight sensor verification.*

# 4. DATA ANALYSIS FOR WEIGHT SENSOR VERIFICATION

This section presents the statistical analysis conducted to determine the decision thresholds of the weight sensor verification mechanism designed in Section 3. Independently of the physical implementation in Section 5 and the testing process in Section 6, the analysis constitutes the theoretical verification infrastructure of the system.

## 4.1. Data Collection

For the weight-sensor-based presence verification mechanism to operate reliably, a statistical weight reference representing the target user population is required. Since sufficient data specific to the 18–22 age group could not be found in national sources, three separate datasets were compiled from the international open data platform Kaggle.

The Gym Members Exercise Dataset (Kaggle, 2024) contains 973 records; the Obesity Classification Dataset (Kaggle, 2023) contains 109 records; and the Medical Cost Personal Datasets (Kaggle, 2018) contain 1,338 records. Since weight values were not directly provided in the last dataset, they were calculated from the BMI = Weight / Height$^2$ formula under the assumption of a fixed height of 170 cm for males and 160 cm for females.

## 4.2. Data Preprocessing

A systematic data cleaning was carried out on all three datasets using the Python programming language with the Pandas and Matplotlib libraries. Only the age, gender, and weight columns were retained; individuals outside the 18–22 age range and values below 40 kg were excluded. The three filtered datasets were merged, resulting in a final sample of 350 individuals.

## 4.3. Statistical Findings

### *Gender Distribution*

Males constitute 51.4% (180 individuals) of the final sample, and females constitute 48.6% (170 individuals) (Figure 7). The nearly equal representation of both genders allows for a balanced interpretation of the analysis findings.

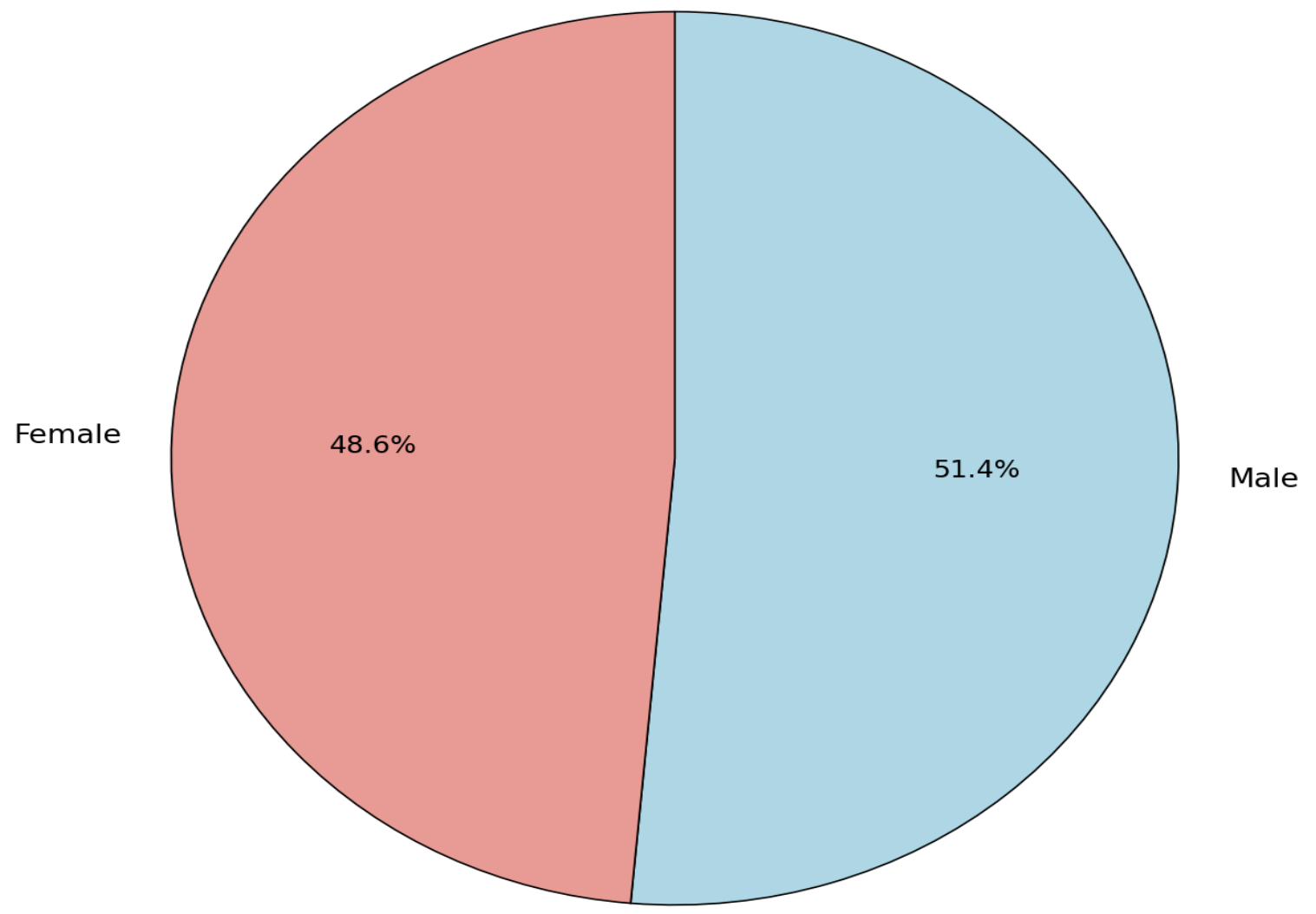

*Figure 7. Gender distribution of the sample (n=350; Male: 51.4%, Female: 48.6%).*

### *Mean Weight Values by Gender and Age*

In all age groups, the mean weight of males is notably higher than that of females (Figure 8). The mean values by age for males are as follows: 85.41 kg at age 18, 82.76 kg at age 19, 94.59 kg at age 20, 85.44 kg at age 21, and 92.52 kg at age 22. For females, these values are 76.96 kg, 69.37 kg, 70.20 kg, 66.86 kg, and 65.11 kg, respectively.

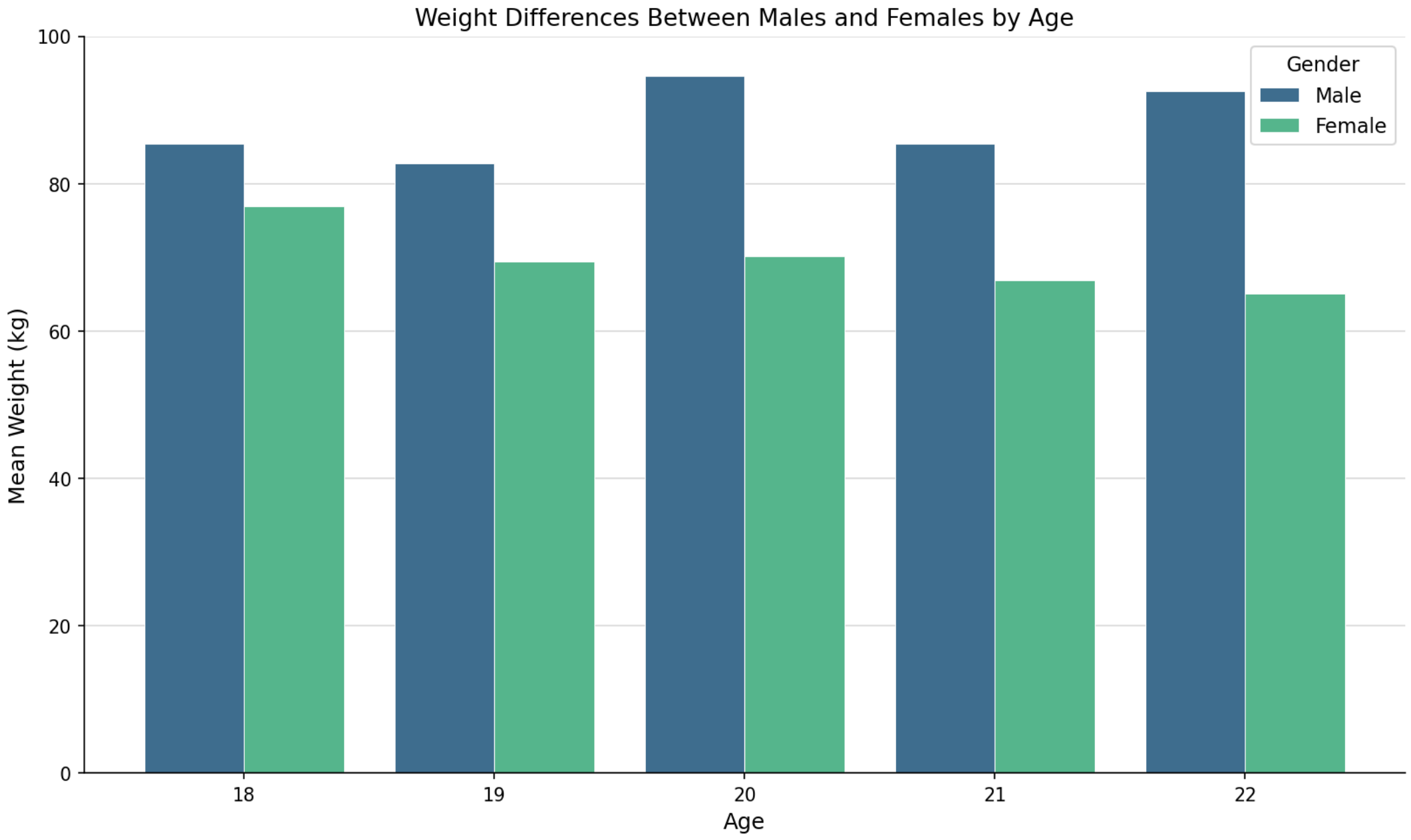

*Figure 8. Weight distributions by age and gender (18–22 age group, n=350).*

Figure 9 shows the mean weight values by age group.

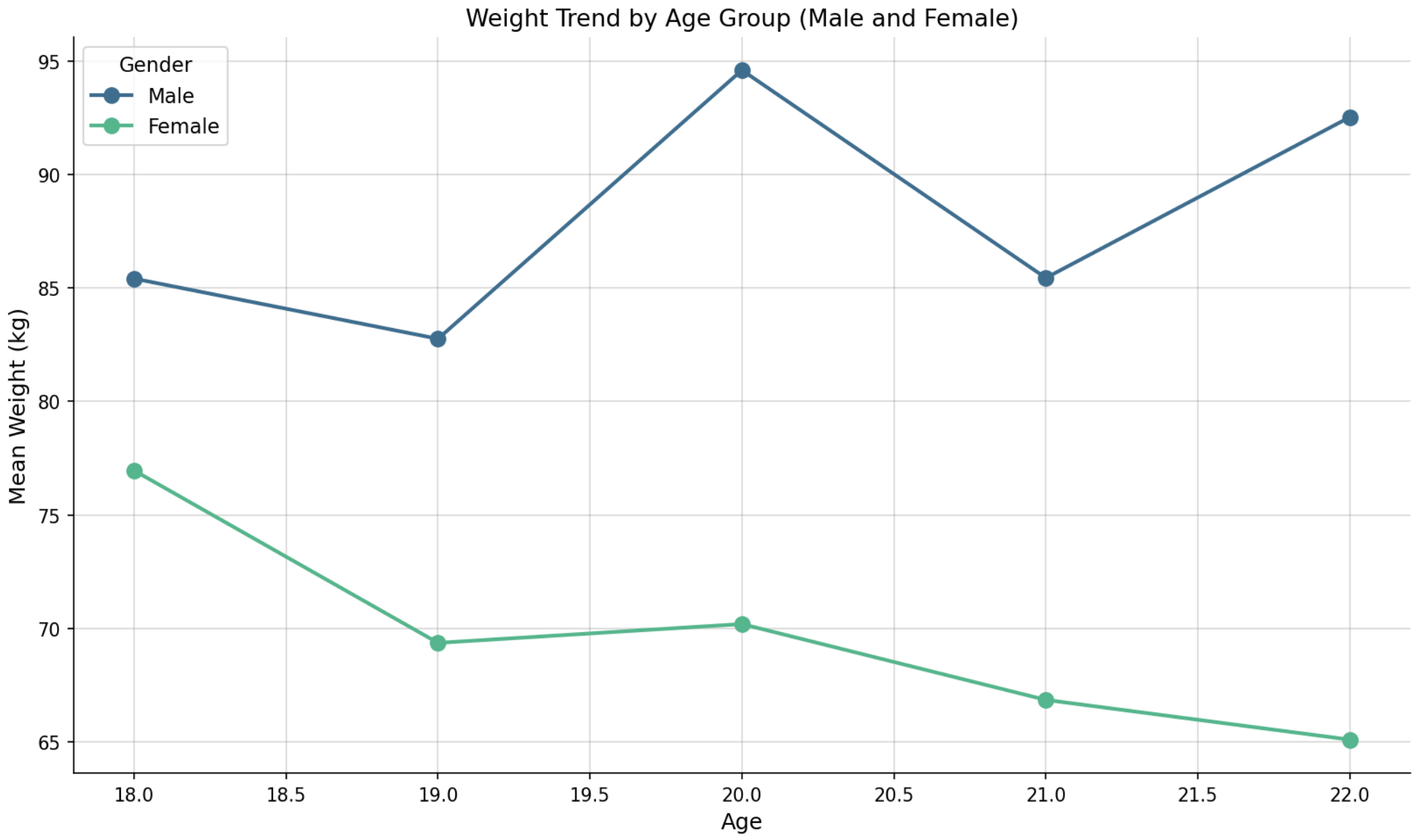

*Figure 9. Mean weight values by age group (18–22 age group, n=350).*

### ***Weight Difference Between Genders***

While the mean weight of males in the 18-year-old group is approximately 11% higher than that of females, this difference rises to 35% in the 20-year-old group. This finding confirms the necessity of defining weight thresholds separately by gender and makes gender-based threshold application in the system mandatory.

### 4.4. Integration of Verification Thresholds into the System

The gender- and age-based mean weight values obtained from the Section 4 analysis form the statistical basis of the “weight pool” mechanism in the system. When a student successfully scans their RFID card, the mean weight value corresponding to the relevant age and gender group is added to the system as a reference. In this way, the expected total weight for registered students in the class can be calculated. The actual total weight measured from the chair sensors is then compared against this expected value, and differences outside the defined tolerance limit are reported to the instructor as anomaly warnings.

At the individual chair level, transient measurements between 10 and 40 kg are evaluated as sitting-down and standing-up transitions and are rechecked three times at one-second intervals. Measurements above 40 kg are considered as active seating. Thus, the system's robustness against sensor noise is increased by using both individual chair-level threshold checking and class-total statistical comparison together.

The weight pool mechanism defined in this study represents the verification logic envisaged for the multi-chair full classroom architecture. However, the prototype presented in this paper has been limited to a single chair at the proof-of-concept level. Therefore, the class-wide total weight comparison has not yet been implemented on full-scale hardware.

The main privacy advantage of this approach is as follows: the only data recorded in the system are the student's age and gender. This information consists of standard demographic data that is already present in institutional record systems and is not considered special category personal data. The instantaneous weight measurement is not recorded in the system; it is used only for one-time comparison.

## 5. IMPLEMENTATION AND PROTOTYPE

This section details the process of physically implementing the system designed in Section 3. Unlike the design decisions in Section 3, this section addresses the actual assembly of the hardware, the wiring connections between components, the software development steps, and the data flow structure.

## 5.1. Hardware Setup

The prototype has a modular hardware architecture centered around a single Arduino UNO board. Four half-bridge load cells were connected to the HX711 ADC module by soldering, and the resulting assembly was wired to the Arduino's digital pins. Figure 10 shows the HX711 module with completed soldering and the connected sensors.

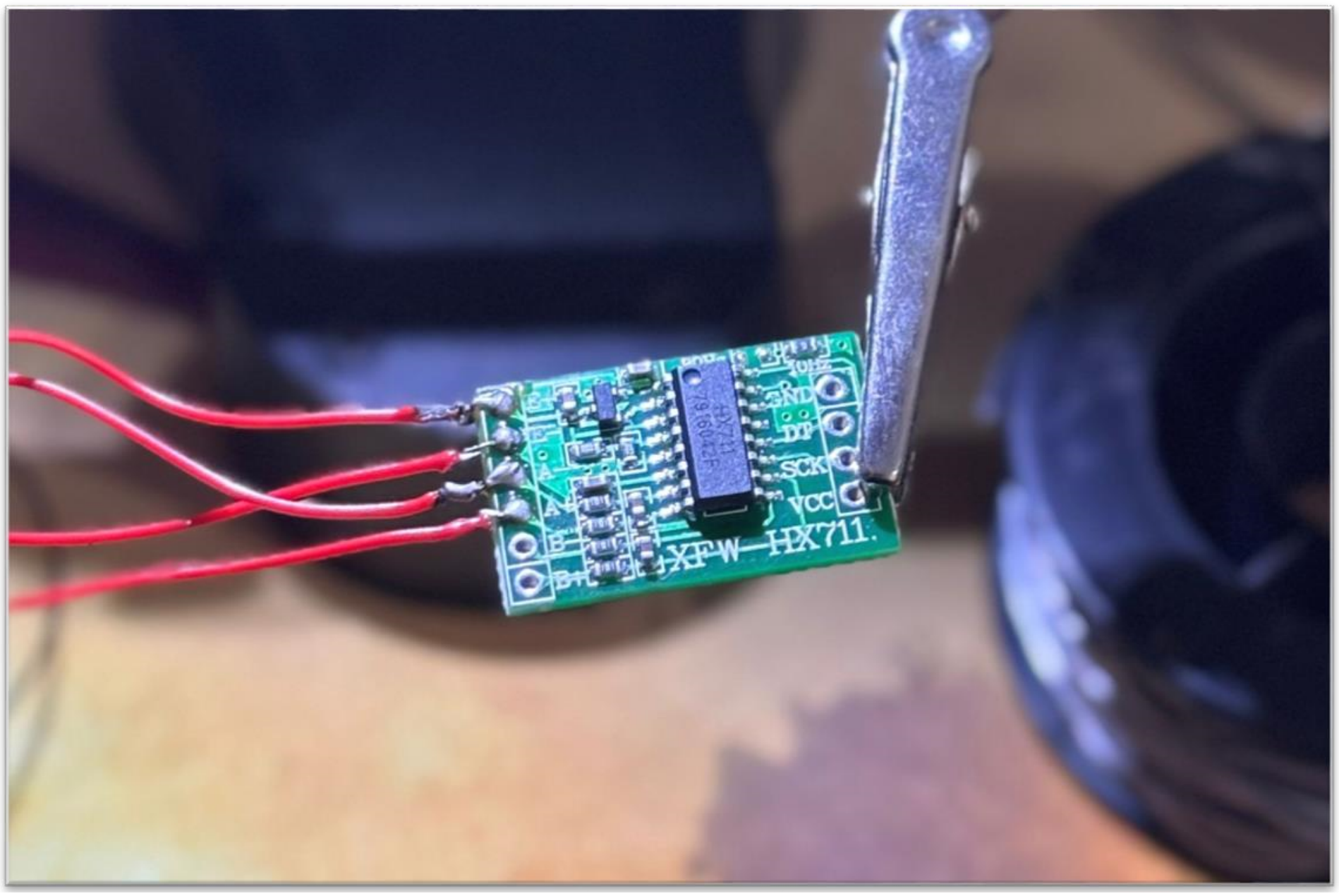

*Figure 10. HX711 ADC module with four soldered weight sensors.*

The sensors were placed at the lower corners of the chair seat board so as to directly bear the student's sitting load.

Color-coded jumper wires were attached to all pins of the RC522 RFID module except IRQ; the SDA pin was connected to the D10 pin of the Arduino UNO, the SCK pin to the D13 pin, the MOSI pin to the D11 pin, and the MISO pin to the D12 pin. Figure 11 shows the RC522 module with completed connections.

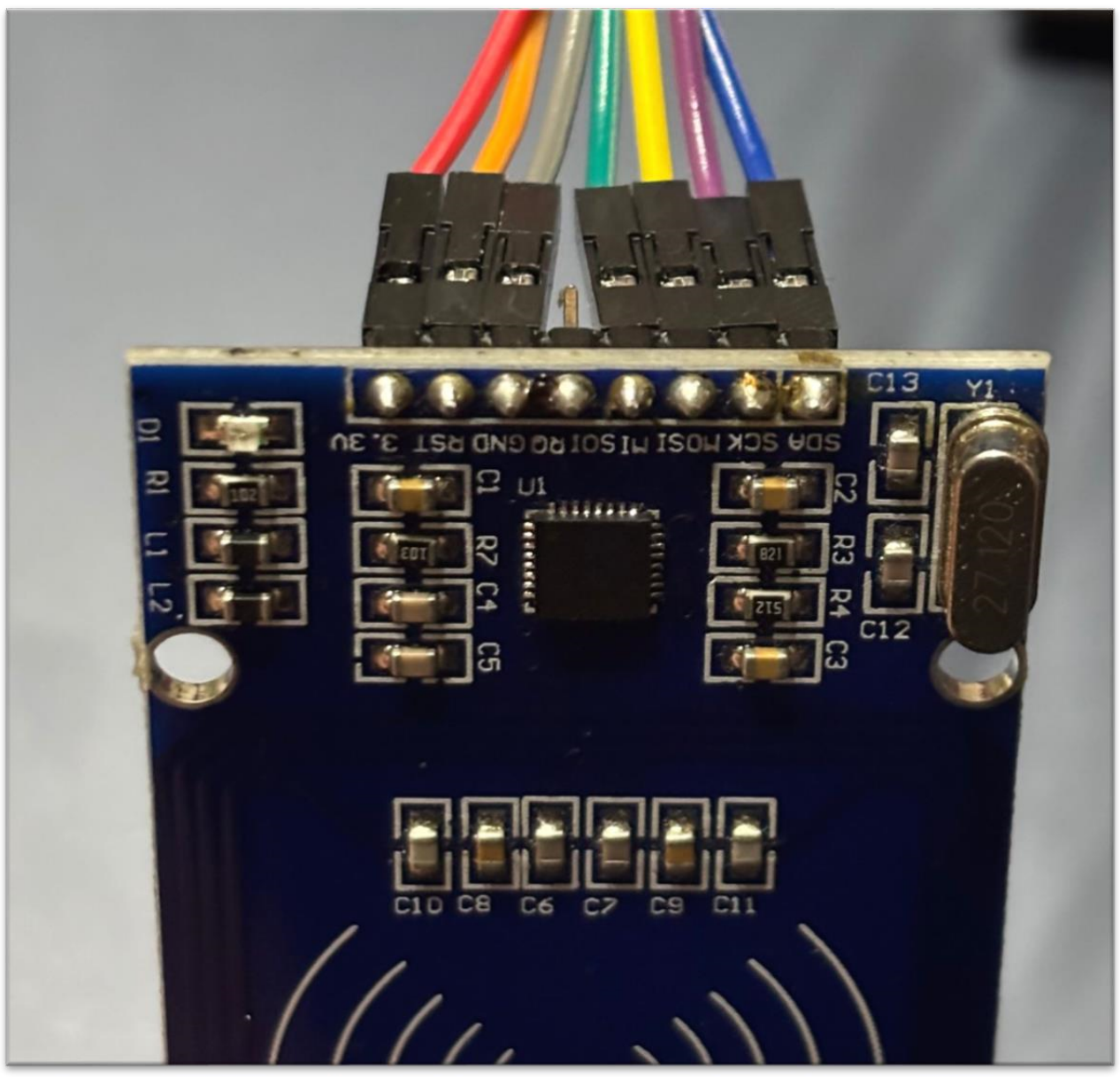

*Figure 11. RC522 RFID module connected with color-coded wires.*

The VCC pin of the HC-06 Bluetooth module was connected to the Arduino's 3.3V pin, the GND pin to the GND pin, the TX pin to the RX (D0) pin, and the RX pin to the TX (D1) pin. The LCD display with the I2C module was connected to the Arduino via only four communication lines. Since all components are powered by the Arduino's 5V output, no external power source is required. Figure 12 presents an overall view of the completed prototype hardware.

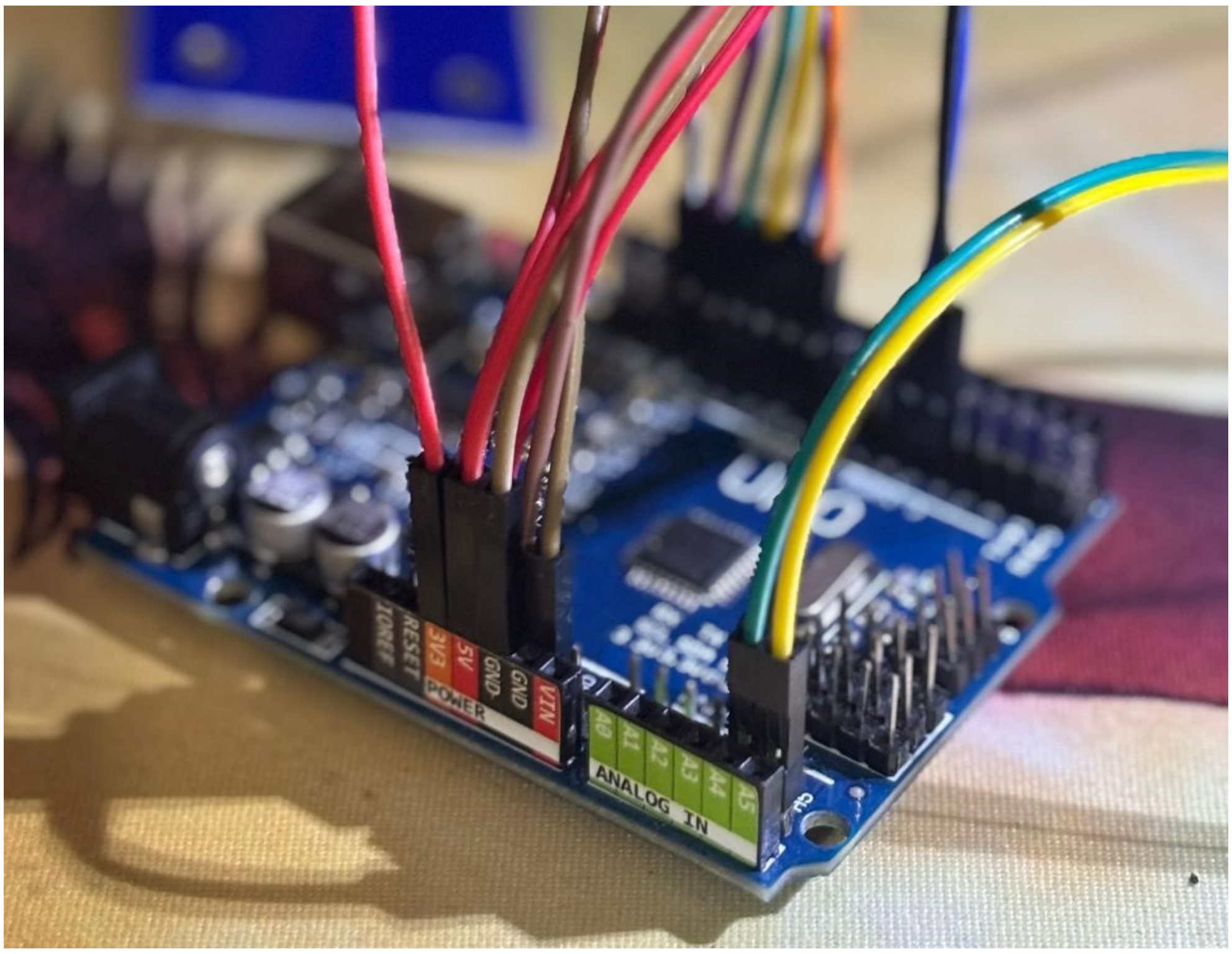

*Figure 12. Overall view of the implemented prototype hardware.*

## 5.2. Software Development and Integration

### *Arduino Embedded Software*

The Arduino embedded software consists of three main modules: RFID reading (including retry logic), weight measurement (converting ADC values to kilograms), and Bluetooth communication. The main loop runs at approximately 10 Hz; this value provides an appropriate balance between response speed and stability.

### *Python GUI Application*

The desktop management interface contains four functional modules: student management, course management, attendance, and reporting. Each student record includes full name, RFID card identifier (UID), age, and gender information. The attendance module processes data coming in over Bluetooth in real time and writes verified entries to the CSV file. Figure 13 shows the main GUI screen.

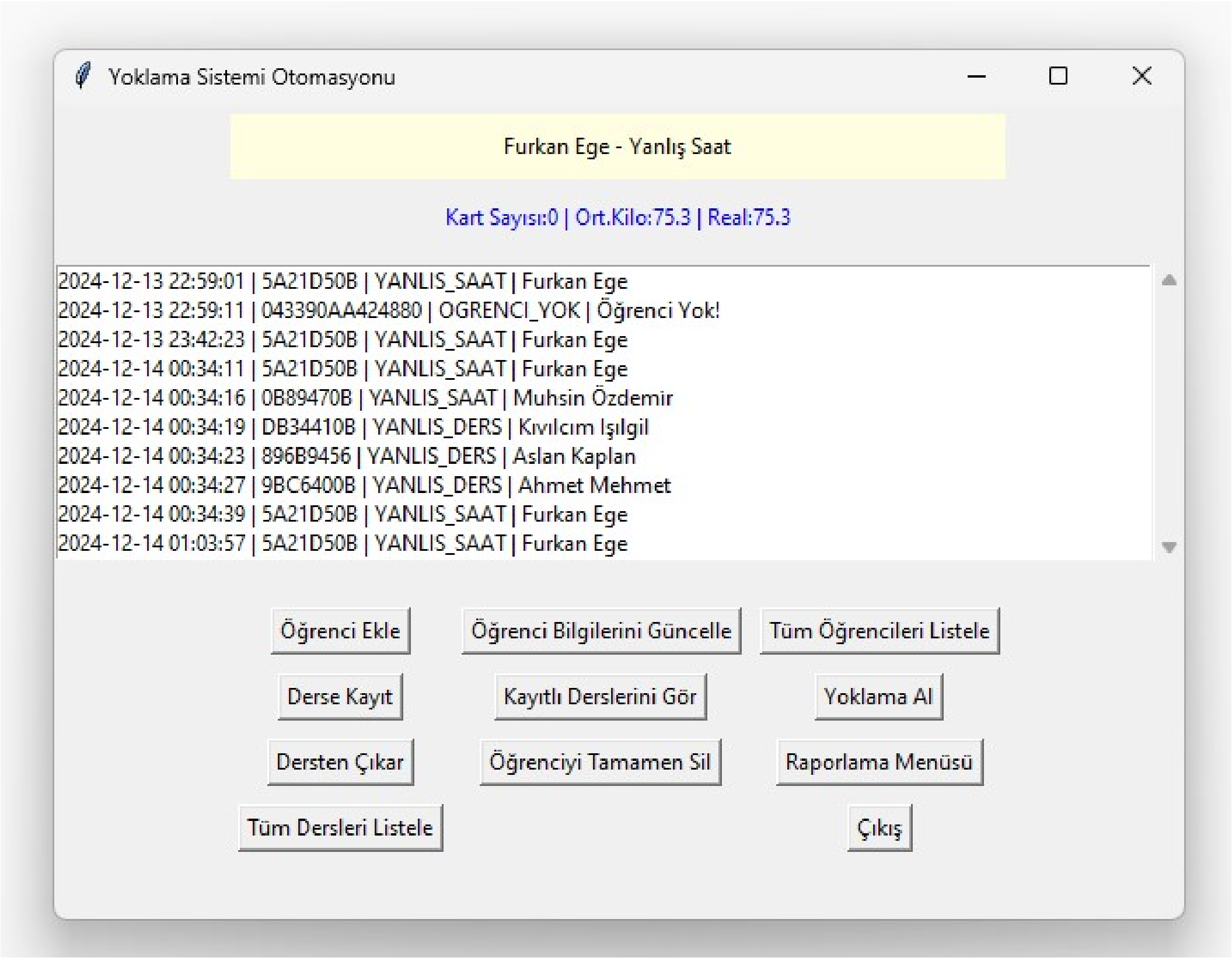

*Figure 13. Python GUI main screen: attendance records, student and course management functions. Note: interface labels are shown in the original Turkish used in the prototype implementation. English equivalents of the key labels are: Öğrenci Ekle = Add Student, Öğrenci Bilgilerini Güncelle = Update Student Information, Tüm Öğrencileri Listele = List All Students, Derse Kayıt = Enroll in Course, Kayıtlı Derslerini Gör = View Enrolled Courses, Yoklama Al = Take Attendance, Dersten Çıkar = Remove from Course, Öğrenciyi Tamamen Sil = Delete Student, Raporlama Menüsü = Reporting Menu, Tüm Dersleri Listele = List All Courses, Çıkış = Exit.*

***Age Information at the Prototype Stage***

At the prototype stage, student age information was simulated in the 18–22 range for the purpose of verifying the functional operation of the system; the current GUI form does not separately include an age field. In real institutional use, this value will be automatically obtained from university information systems and directly matched against the statistical reference table in Section 4.

This design decision, while limiting the current scope of the system at the proof-of-concept level, does not affect the fundamental mechanism of the age-based verification logic.

***Data Storage Structure***

The system adopts a CSV-based storage architecture that does not require institutional database software. Student records, course information, and timestamped attendance entries are kept in separate files; this structure facilitates import/export compatibility with existing university information systems.

## 5.3. System Workflow

The attendance-taking cycle operates as follows: the instructor starts an attendance session by selecting the relevant course through the GUI. The system allows an attendance record to be created only for the course matching the current time slot.

When a student scans their RFID card, the system first sequentially checks the conditions defined in Section 3.4. If the card is registered in the system, the student is enrolled in the relevant course, and the scanning time matches the course time, the process proceeds to the weight verification stage. At this stage, the measured value is evaluated within the framework of the age- and gender-based statistical reference logic described in Section 4.

For measurements falling within the tolerance range, attendance is confirmed and the transaction is recorded in the CSV file along with its timestamp. Thus, the system forms the attendance decision by sequentially combining identity verification with physical presence verification. The entire workflow is completed within seconds and has minimal impact on class time.

# 6. TESTING PROCESS AND FINDINGS

This section reports the testing process carried out on the prototype implemented in Section 5 and the findings obtained. Unlike Section 5, which addresses the assembly and development process, this section focuses on the test results documenting how the system operates and how various issues were resolved.

## 6.1. Testing Approach

A two-stage testing strategy was adopted to evaluate the system's functional correctness and integrated performance. In the first stage, each module was tested independently; in the second stage, the modules were operated together and end-to-end system behavior was tested in an environment close to real classroom conditions. Table 1 summarizes the tested scenarios, expected behaviors, and observed results. These tests were conducted as qualitative verification tests aimed at evaluating whether the prototype exhibits the expected functional behavior. Systematic quantitative success rates and error metrics are not reported within the scope of this study.

*Table 1. Test scenarios, expected behaviors, and observed results.*

| **Test Scenario** | **Expected Behavior** | **Observed Result** |
|---|---|---|
| Registered card + correct course time | Display of the “Welcome” message on the LCD screen and creation of an attendance record | Expected behavior observed |
| Unregistered card | Display of the “Student Not Found” warning on the LCD screen and no record created | Expected behavior observed |
| Registered card + incorrect course time | Display of the “Wrong Time” warning on the LCD screen and no record created | Expected behavior observed |
| Registered card + unregistered course | Display of the “Wrong Course” warning on the LCD screen and no record created | Expected behavior observed |
| Rescanning the same card | Consistent maintenance of authentication and preservation of memory stability | Expected behavior observed |
| Different reference weights | After calibration, measurements match kilogram values consistently | Expected behavior observed |
| With Arduino IDE monitor open | Detection of the GUI's inability to access the COM port and establishment of appropriate usage protocol | Protocol established |
| Long-duration operation | Stable continuation of memory, port, and file operations | Expected behavior observed |

## 6.2. Module-Level Tests

### *RFID Reading Module*

The RC522 module was tested with both registered and unregistered cards. For registered cards, the system consistently identified the card ID correctly; after the initial screen shown in Figure 14, the confirmation message in Figure 15 was displayed on the LCD and data was transmitted via Bluetooth.

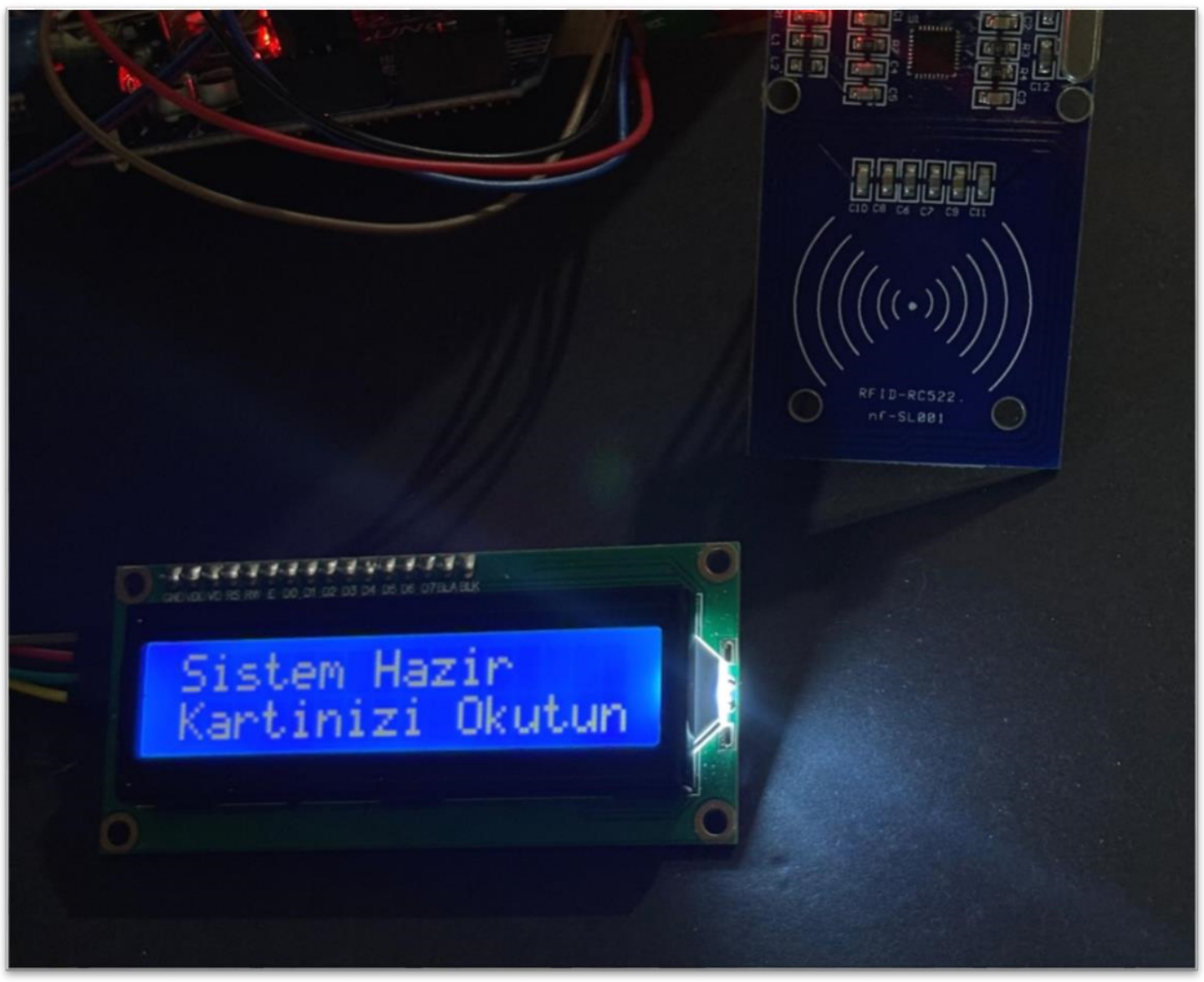

*Figure 14. LCD display at system startup: "Sistem Hazir, Kartinizi Okutun" (English: "System Ready, Scan Your Card"). Note: the LCD displays Turkish text as programmed in the prototype firmware.*

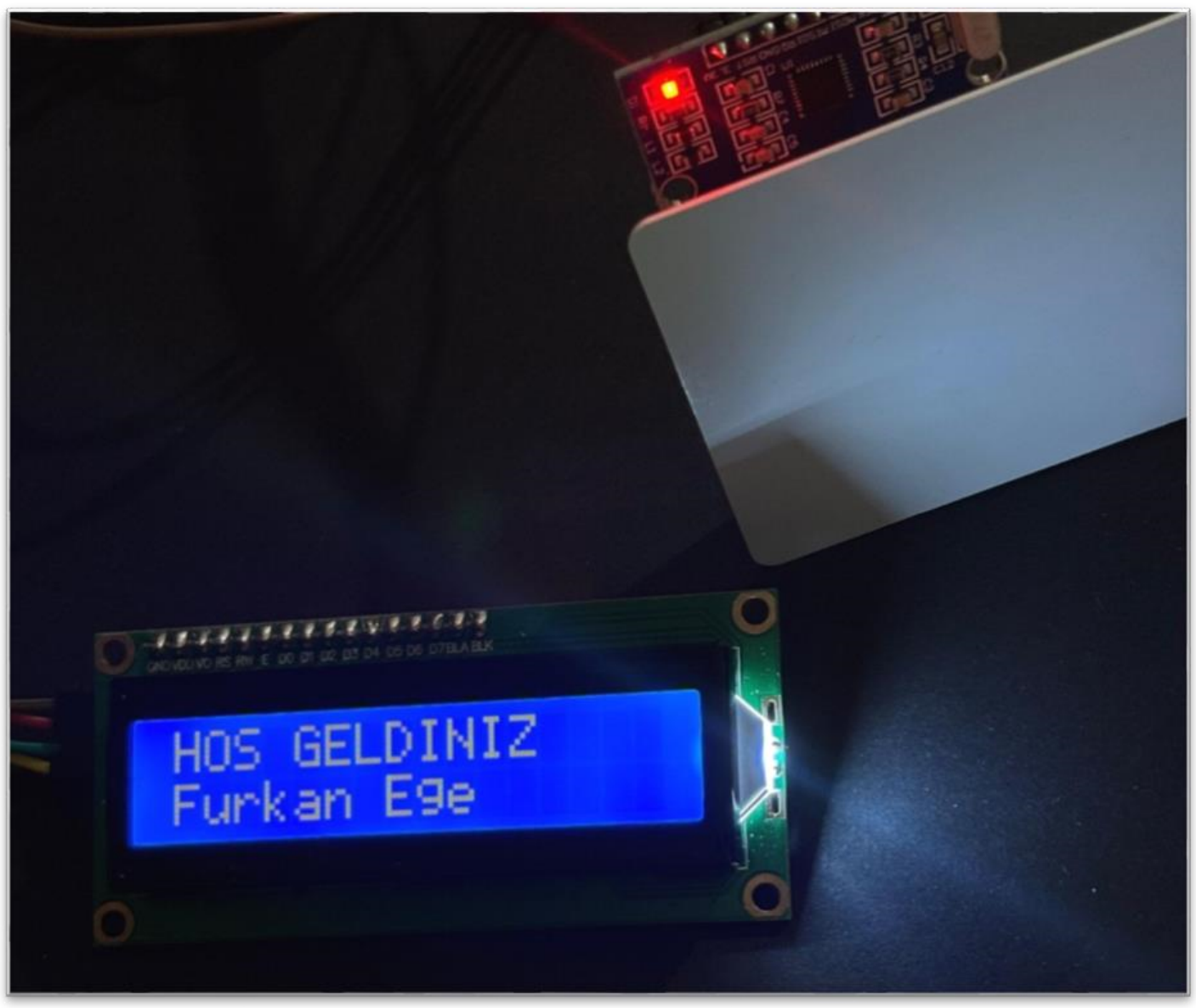

*Figure 15. LCD display when a registered card is scanned: "HOS GELDINIZ" (English: "Welcome") followed by the student's name. Note: the LCD displays Turkish text as programmed in the prototype firmware.*

For unregistered cards, the "Student Not Found" warning shown in Figure 16 was displayed on the screen and no attendance record was created. It was observed that card reading was occasionally not completed on the first attempt; this issue was resolved by the retry logic added to the software.

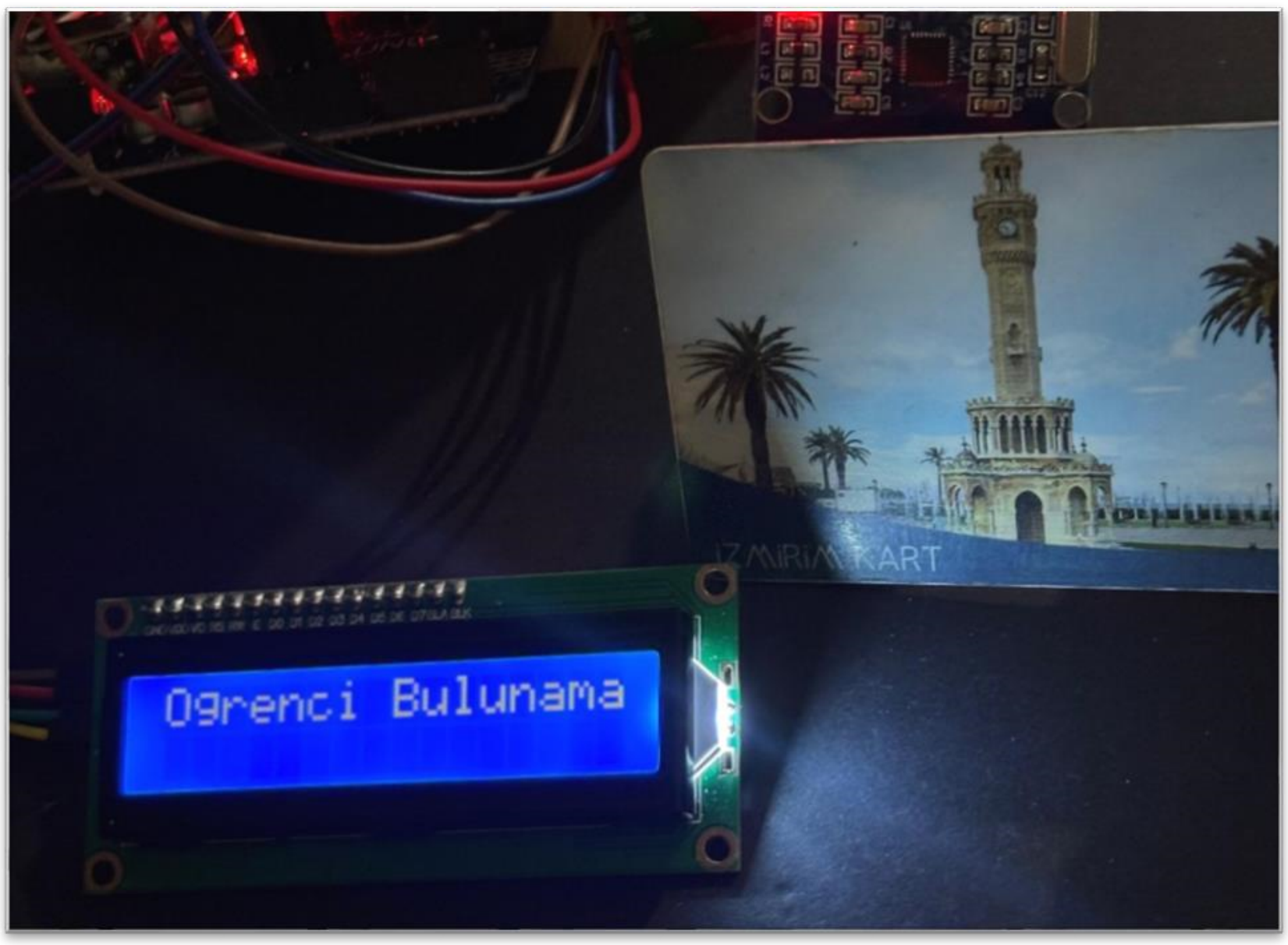

*Figure 16. LCD display when an unregistered card is scanned: "Ogrenci Bulunama" (English: "Student Not Found"). Note: the LCD displays Turkish text as programmed in the prototype firmware.*

### *Weight Sensor Module*

The load cells were systematically tested after being calibrated with known reference weights. Consistent results were obtained in the conversion of ADC values to kilograms. It was observed that ambient vibrations caused instability in the measurements; this issue was resolved by the tolerance-based filtering algorithm applied at the software layer.

### *Bluetooth Communication Module*

The HC-06 module was tested in terms of the data flow between the Arduino and the Python GUI. It was confirmed that data transmission generally occurred reliably; however, occasional delays were observed.

By optimizing the data transmission frequency, delays were reduced to an acceptable level. Additionally, it was found that the Python GUI could not access the COM port while the Arduino IDE Serial Port monitor was open; it was concluded that tests should be conducted only while the GUI is running and the Serial Port monitor is closed.

***Python GUI Module***

All functional components of the interface were tested separately. Error scenarios, including time/course matching checks, scanning of cards not registered for the course, and attempts outside the course time, were systematically tested.

In scanning attempts carried out outside course hours, the wrong time warning shown in Figure 17 was successfully triggered. In all scenarios, the system exhibited the expected behavior.

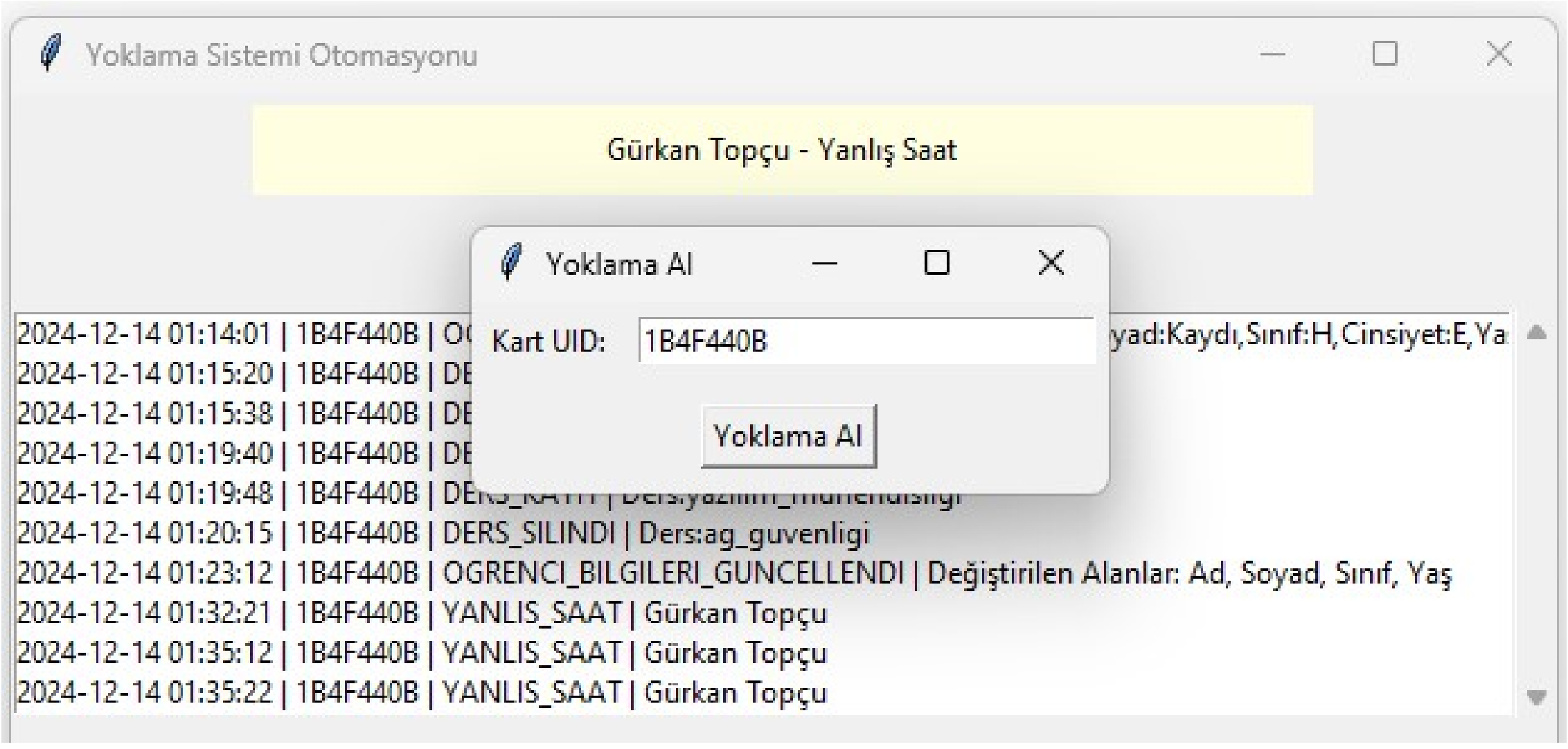

*Figure 17. "Yanlis Saat" warning (English: "Wrong Time") for scanning outside course hours. Note: GUI and LCD labels are shown in the original Turkish used in the prototype. Key labels visible in the figure: Yoklama Sistemi Otomasyonu = Attendance System Automation, Yoklama Al = Take Attendance, Kart UID = Card UID, YANLIS_SAAT = WRONG_TIME, DERS_KAYIT = COURSE_ENROLLMENT, DERS_SILINDI = COURSE_DELETED, OGRENCI_BILGILERI_GUNCELLENDI = STUDENT_INFORMATION_UPDATED, Degistirilen Alanlar = Modified Fields, Ad, Soyad, Sinif, Yas = Name, Surname, Class, Age.*

## 6.3. Integrated System Test

After the module tests were completed, the system was tested as a whole under conditions close to the actual classroom setting. The integrated tests showed that four fundamental functions were performed smoothly: successful recording of valid card scans, rejection of unauthorized card attempts, warnings for scans outside course hours, and consistent results produced by cross-checking of weight and RFID verification. Figure 18 shows the attendance records obtained on the GUI during the real operation of the system.

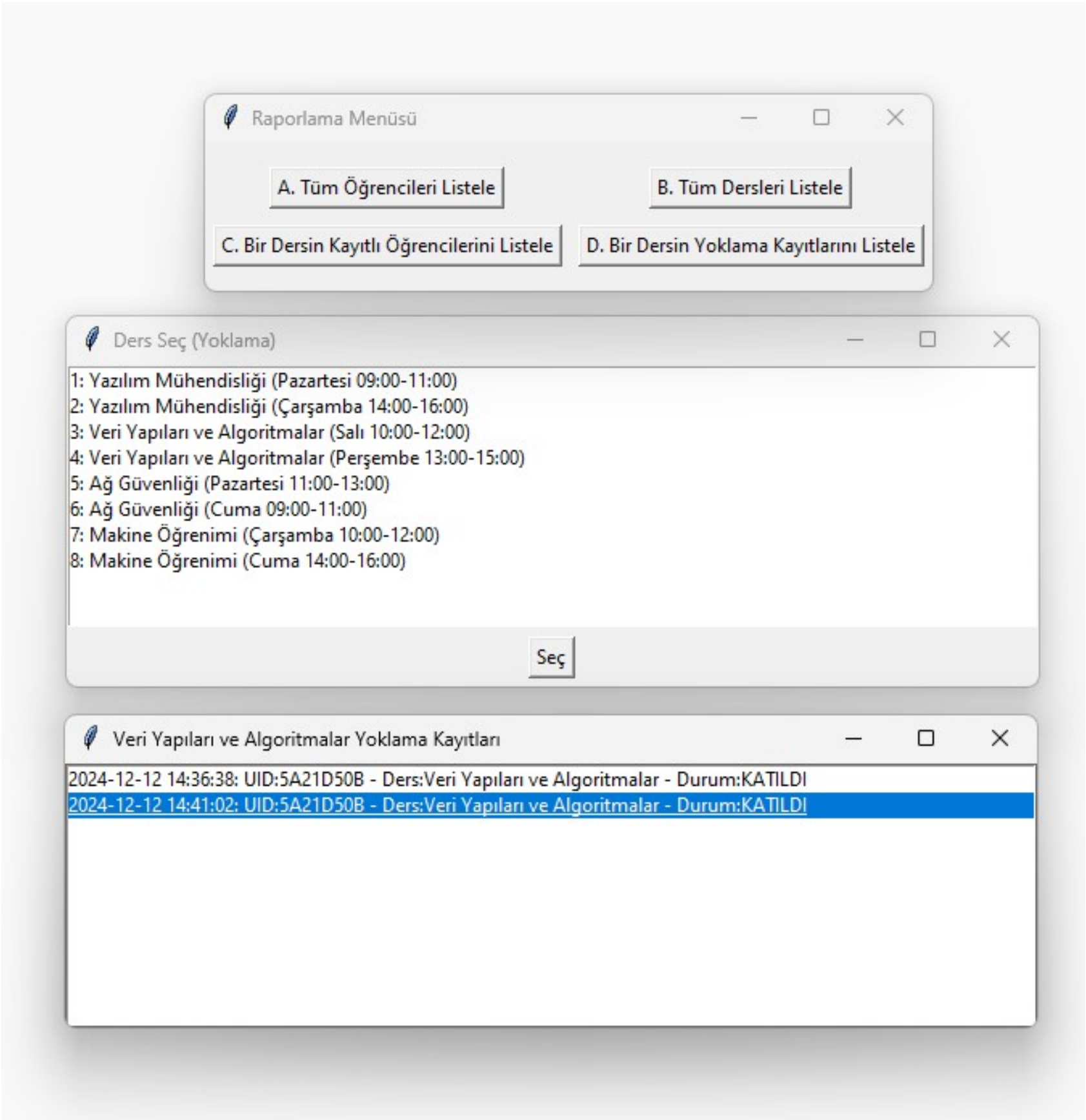

*Figure 18. Attendance records obtained on the GUI during real operation. Note: GUI labels are shown in the original Turkish used in the prototype. Key labels visible in the figure: Raporlama Menüsü = Reporting Menu, Tüm Öğrencileri Listele = List All Students, Tüm Dersleri Listele = List All Courses, Bir Dersin Kayıtlı Öğrencilerini Listele = List Students Enrolled in a Course, Bir Dersin Yoklama Kayıtlarını Listele = List Attendance Records of a Course, Ders Seç (Yoklama) = Select Course (Attendance), Seç = Select, Yazılım Mühendisliği = Software Engineering, Veri Yapıları ve Algoritmalar = Data Structures and Algorithms, Ağ Güvenliği = Network Security, Makine Öğrenimi = Machine Learning, Pazartesi = Monday, Salı = Tuesday, Çarşamba = Wednesday, Perşembe = Thursday, Cuma = Friday, Yoklama Kayıtları = Attendance Records, Ders = Course, Durum = Status, KATILDI = ATTENDED.*

The system's memory usage and file read/write functions remained stable throughout the testing.

### 6.4. Encountered Limitations

The prototype supports only a single chair at a time; extending it to a multi-chair environment requires additional hardware and software. The weight verification mechanism carries the possibility of generating false warnings for individuals falling outside the statistical range in Section 4. System performance has not been comprehensively tested under conditions of vibration, temperature variation, and electromagnetic interference that could be encountered in an actual university classroom.

# 7. DISCUSSION

## 7.1. Comparison of the Proposed System with Existing Approaches

The system developed in this study proposes an architecture that simultaneously addresses two fundamental problems in the electronic attendance literature: the requirement for biometric data and the proxy attendance gap.

Biometric-based systems report high accuracy rates. Soewito et al. (2016) reported 95% accuracy, and Dey et al. (2014) reported 94.2% accuracy. However, all such systems rely on the central processing or storage of the individual's permanent physical characteristics (Kindt, 2013). The proposed system, on the other hand, does not store any biometric data; in this respect, it significantly reduces the data protection obligations in the context of GDPR, KVKK, and FERPA.

Systems using only RFID do not incorporate any built-in safeguard against the proxy attendance problem when the card is transferred (Duroc, 2022). The proposed system closes this gap through weight sensor integration. Although hybrid approaches combining RFID with biometric verification largely solve the proxy attendance problem, they do not eliminate the obligation to store biometric data. In terms of cost, the total cost of passive RFID tags and load cells is significantly below the installation cost of fingerprint readers and iris scanners (Want, 2006).

## 7.2. Evaluation of Limitations

The weight pool mechanism defined in this study represents the verification logic envisaged for the multi-chair full classroom architecture. However, the prototype presented in this paper has been limited to a single chair at the proof-of-concept level. For this reason, the class-wide total weight comparison has not yet been implemented on full-scale hardware.

The extent to which the 350-individual sample drawn from Kaggle sources represents different geographical and demographic groups remains uncertain. In part of the sample, weight values were derived through BMI under a fixed-height assumption. This indicates that the reference distribution has a partially synthetic character. Furthermore, the possibility of generating false warnings for extremely underweight or overweight students constitutes a usability issue that must be carefully evaluated. In addition, student age information was simulated at the prototype stage. In actual institutional use, this data must be obtained automatically from university information systems. Finally, the long-term field performance of the system has not yet been comprehensively and quantitatively evaluated.

### 7.3. Future Work

Scaling the system in a multi-chair environment, providing integration with an institutional record system using real student age data, and establishing a quantitative validation framework constitute the priority agenda items for future research. The addition of a mobile application layer will enable real-time access to attendance information by students. The development of machine-learning-based absence prediction models may lay the groundwork for deeper integration of the system into decision support processes.

## 8. CONCLUSION

In this study, a two-layer IoT architecture that simultaneously addresses two fundamental problems in the field of electronic attendance — the requirement for biometric data and the physical presence verification gap in RFID-based systems — has been developed and tested at the prototype level. The original contribution of the system lies in separating the tasks of identity verification (RFID) and physical presence verification (weight sensor) into two independent layers. Thanks to this architectural choice, proxy attendance occurring when the card is transferred can be detected through a statistical reference framework without having to store biometric data.

The four main contributions of the study can be summarized as follows. First, a privacy-focused attendance architecture that does not store any biometric data has been put forward; the instantaneous weight measurement is not recorded but is subjected only to a one-time comparison, and this structure structurally reduces the data protection obligations in the context of GDPR, KVKK, and FERPA. Second, a novel proxy attendance detection mechanism that integrates RFID identity verification with weight-sensor-based physical presence verification and is built on the weight pool mechanism has been designed. Third, the weight distribution by age and gender has been systematically analyzed on a 350-individual sample compiled from three Kaggle datasets; this analysis provides the basis for both individual chair threshold logic and class-wide weight pool calculation (Kaggle, 2018; Kaggle, 2023; Kaggle, 2024). Fourth, the prototype was tested in an environment close to real classroom conditions; the integrated operation of the RFID reading, weight verification, Bluetooth communication, and GUI modules exhibiting the expected functions was observed in eight different qualitative test scenarios (Table 1).

The main limitations of the study are as follows: the prototype was implemented at a single-chair scale, age information was simulated at the proof-of-concept stage, and systematic quantitative validation metrics have not yet been established. Nevertheless, this study has addressed, through a concrete prototype, a design gap that existing systems in the literature have not been able to close; it has shown that a reliable classroom attendance system can be implemented without depending on biometric authentication. Multi-chair scaling, institutional integration with real age data, and the establishment of a quantitative validation framework are the priority goals of future research that would naturally extend this study.

# RFID Tabanlı ve Biyometrik Kimliklendirmesiz Sınıf Yoklama Sistemi: Ağırlık Sensörü Entegrasyonuyla Vekâleten Yoklama Tespiti

**Furkan EGE[1], Muhsin ÖZDEMİR[1]**

[1] *Yönetim Bilişim Sistemleri Anabilim Dalı, Sosyal Bilimler Enstitüsü, Aydın Adnan Menderes Üniversitesi, Aydın, Türkiye*

## ÖZET

Eğitim kurumlarında derse devam takibi, geleneksel yöntemlerle yürütüldüğünde ders süresini tüketen ve akademik dürüstlüğü tehdit eden yapısal sorunlara yol açmaktadır. İlk ve ortaöğretimde 3–6 dakika, yükseköğretimde 10 dakikayı aşabilen yoklama süreleri ve "başkasının yerine imza atma" biçiminde somutlaşan vekâleten yoklama problemi, elektronik yoklama sistemlerine duyulan ihtiyacı ortaya koymaktadır. Mevcut elektronik çözümlerin çoğu biyometrik kimlik doğrulamaya başvurmakta; bu durum GDPR, KVKK ve FERPA kapsamında hukuki ve etik riskler doğurmaktadır. Yalnızca RFID kullanan sistemler ise kart devri yoluyla gerçekleştirilen vekâleten yoklamaya karşı yapısal güvence sunamamaktadır.

Bu çalışmada, iki eksikliği eş zamanlı gideren, biyometrik veriden bağımsız bir IoT yoklama sistemi önerilmektedir. Sistem prototipi RFID modülü, RFID kartları, ağırlık sensörleri, Bluetooth modülü ve Arduino UNO mikrodenetleyicisinden oluşmaktadır. Öğrencinin RFID kartını okutmasının ardından ağırlık sensöründen alınan ölçüm, Kaggle'dan derlenen üç veri setine dayalı 350 bireylik (18–22 yaş) istatistiksel referans aralığıyla karşılaştırılmakta; süreçte biyometrik veri kaydedilmemektedir. Python tabanlı GUI, Bluetooth üzerinden öğrenci yönetimi, ders takibi ve CSV tabanlı raporlama işlevlerini yürütmektedir.

Gerçek sınıf koşullarına yakın ortamda yürütülen nitel testler, RFID okuma, ağırlık doğrulama, Bluetooth iletişimi ve GUI modüllerinin bütünleşik olarak beklenen işlevleri sergilediğini göstermiştir. Önerilen sistem, biyometrik veri depolamadan vekâleten yoklamayı azaltmayı hedefleyen, düşük maliyetli ve yeniden üretilebilir bir çözüm sunmaktadır.

# 1. GİRİŞ

Eğitim, bireyin davranışında kendi yaşantısı yoluyla ve kasıtlı olarak istendik değişimler meydana getirme süreci olup eğitim kurumları, bu sürecin gerçekleştiği ve öğrencilere bilgi, beceri ile tutum kazandırılmaya çalışılan kurumlardır (Ertürk, 1997: 12). Bu sürecin vazgeçilmez bir parçası olan ders devam takibi, öğrencilerin akademik başarısıyla doğrudan ilişkili olmakla birlikte yönetimsel açıdan da eğitim kurumlarının sürdürülebilir işleyişini destekleyen kritik bir bileşendir. Ne var ki öğrenci devam durumunun kayıt altına alınmasında yaygın biçimde başvurulan imzaya veya sözlü yoklamaya dayalı geleneksel yöntemler, değerli ders süresini tüketen ve ciddi güvenilirlik sorunları barındıran bir yapı sergilemektedir.

Günümüzde ilk ve ortaöğretim kurumlarında uygulanan sözel yoklama yöntemi, öğretmenin öğrenci isimlerini tek tek seslendirmesi ve öğrencilerin yanıt vermesiyle yürütülmektedir. Bu süreç, sınıfın büyüklüğüne bağlı olarak 3 ila 6 dakika sürmektedir. Yükseköğretim kurumlarında ise imzalı devam çizelgesi yöntemi yaygın olarak uygulanmakta; öğrenci sayısının yüksekliği nedeniyle bu süre 10 dakikayı aşabilmektedir (Sezdi ve Tüysüz, 2018). Söz konusu zaman kaybının yanı sıra imzalı yoklama yöntemi, "başkasının yerine imza atma" şeklinde somutlaşan vekâleten yoklama sorununu da beraberinde getirmektedir.

Söz konusu sorunlara çözüm üretmek amacıyla çok sayıda elektronik yoklama sistemi önerilmiştir. Parmak izi tanıma, yüz tanıma, iris tarama, ses biyometrisi, NFC, Bluetooth, Wi-Fi ve QR kod gibi farklı teknolojilere dayanan bu sistemler, yoklama sürecini otomatikleştirerek hem zaman tasarrufu sağlamayı hem de vekâleten yoklamayı engellemeyi hedeflemektedir (Ishaq ve Bibi, 2023). Ancak bu yaklaşımların büyük çoğunluğu biyometrik veri toplamayı zorunlu kılmaktadır. Bu veriler, olası bir veri ihlali durumunda kalıcı zararlar doğurabilmekte (Kindt, 2013) ve GDPR (European Union, 2016), T.C. KVKK ile ABD FERPA (U.S. Department of Education, 1974) kapsamında özel nitelikli kişisel veri ya da korunan eğitim kaydı olarak düzenlenmektedir; bu durum söz konusu sistemlerin eğitim kurumlarındaki uygulanabilirliğini ciddi ölçüde kısıtlamaktadır.

Öte yandan, yalnızca Radyo Frekansıyla Kimliklendirme (RFID) teknolojisine dayanan sistemler de kendi başlarına yeterli doğrulama güvencesi sunamamaktadır (Duroc, 2022). Öğrencinin RFID kartını başka bir kişiye emanet etmesi, bu sistemlerin en temel zaafiyetini oluşturmaktadır.

Bu çalışmada, biyometrik veri gerektirme ile vekâleten yoklamaya karşı yetersiz güvence gibi iki temel eksikliği eş zamanlı olarak gideren bütünleşik bir yoklama sistemi önerilmektedir. Geliştirilen sistem; RFID teknolojisini, ağırlık sensörlerini, Bluetooth modülünü ve Arduino UNO mikrodenetleyicisini bir araya getirmektedir. Ağırlık sensöründen alınan anlık ölçüm, 18–22 yaş arası 350 bireylik istatistiksel veri setlerinden türetilen aralıklarla karşılaştırılmaktadır; bu yaklaşım biyometrik veri depolamayı gerektirmediğinden GDPR, KVKK ve FERPA bağlamındaki veri koruma yükümlülüklerini önemli ölçüde azaltmaktadır.

Bu makalenin başlıca katkıları şöyle özetlenebilir: (i) biyometrik veri gerektirmeyen, gizlilik odaklı bir yoklama mimarisi; (ii) RFID kimlik doğrulamasını ağırlık sensörü tabanlı fiziksel varlık denetimiyle bütünleştiren özgün bir vekâleten yoklama tespit mekanizması; (iii) gerçek sınıf koşullarına yakın bir ortamda test edilmiş, düşük maliyetli ve yeniden üretilebilir bir IoT prototipi; ve (iv) 18–22 yaş grubu ağırlık dağılımına ilişkin sistematik veri analizi.

Makalenin geri kalanı şu yapıda düzenlenmiştir: İkinci bölümde literatür incelenmekte ve araştırma boşluğu ortaya konulmaktadır. Üçüncü bölümde sistemin tasarım kararları, bileşen seçimleri ve mimari yapısı tanıtılmaktadır. Dördüncü bölümde ağırlık doğrulama eşiklerinin istatistiksel temeli sunulmaktadır. Beşinci bölümde prototip fiziksel montajı ve yazılım geliştirme süreci ayrıntılandırılmakta, altıncı bölümde modül ve bütünleşik sistem testlerinin sonuçları aktarılmaktadır. Yedinci bölümde önerilen sistem tartışılmakta; sekizinci bölümde sonuçlar ve gelecek araştırma yönleri sunulmaktadır.

## 2. İLGİLİ ÇALIŞMALAR

Elektronik yoklama sistemleri üzerine yürütülen araştırmalar, kullanılan kimlik doğrulama teknolojisine göre üç ana başlık altında ele alınabilir: biyometrik tabanlı sistemler, biyometrik olmayan sistemler ve hibrit yaklaşımlar. Bu bölümde söz konusu kategorilerdeki temsil edici çalışmalar ele alınmakta ve mevcut araştırma boşluğu ortaya konulmaktadır.

## 2.1. Biyometrik Tabanlı Sistemler

Biyometrik kimlik doğrulama, akademik yazındaki elektronik yoklama çalışmalarının en geniş kategorisini oluşturmaktadır. Nawaz vd. (2009), parmak izi sensörlerini sınıf girişlerine yerleştirerek öğrenci yoklamalarını elektronik ortamda kaydeden ilk kapsamlı sistemlerden birini geliştirmiştir. Koçak ve Yeleç (2017), biyometrik parmak izi okuyucusuna dayanan ve kullanıcı testlerinde yüksek memnuniyet elde eden bir sistem sunmuştur. Soewito vd. (2016), parmak izi doğrulamasını ses tanımayla birleştiren akıllı telefon tabanlı bir sistem önermiş; parmak izi doğrulamasında %95 eşleşme oranına ulaşmıştır. Kamelia vd. (2018) ise parmak izi modülünü GPS ile bütünleştirerek gerçek zamanlı konum ve yoklama izlemesi gerçekleştirmiş; doğrulama işlemi ortalama 1,39 saniyede tamamlanmıştır.

Yüz tanıma teknolojisine dayanan çalışmalar da bu alanda önemli bir yer tutmaktadır. Sawhney vd. (2019), PCA ve CNN algoritmalarını kullanarak gerçek zamanlı yüz tanıma temelli otomatik yoklama sistemi geliştirmiştir. Lukas vd. (2016), DWT ve DCT yöntemleriyle yüz özelliklerini çıkarmış ve 148 tanıma işleminde %82 başarı oranı elde etmiştir. Turan ve Doğan (2024), yüz tanıma tabanlı güncel bir öğrenci takip sistemi geliştirmiştir. Dey vd. (2014), i-vektör tabanlı konuşmacı tanıma modellemesi kullanan bir ses biyometrisi platformu kurmuş ve iki aylık testte %94,2 doğruluk oranı raporlamıştır. Okokpujie vd. (2017) da web arayüzüyle entegre bir iris tarama sistemi sunmuştur.

Walia ve Jain (2016), parmak izi tabanlı biyometrik yoklama sistemlerinin genel durumunu değerlendiren kapsamlı bir derleme çalışmasıyla bu sistemlerin temel sınırlılığına dikkat çekmiştir: değiştirilemez kişisel özelliklerin merkezi veri tabanlarında depolanması. GDPR, KVKK ve FERPA gibi mevzuatlar bu verileri özel nitelikli kişisel veri ya da korunan eğitim kaydı olarak düzenlemekte; dolayısıyla söz konusu sistemlerin eğitim kurumlarında konuşlandırılması ciddi yasal yükümlülükler doğurmaktadır (Kindt, 2013).

## 2.2. Biyometrik Olmayan Sistemler

Biyometrik olmayan sistemler, kimlik doğrulama için optik, kablosuz sinyal ve RFID teknolojilerine başvurmaktadır. Optik ve yakın alan iletişimine dayanan çalışmalar arasında Masalha ve Hirzallah (2014), akıllı telefon taraması ve çok faktörlü kimlik doğrulamayla yoklamanın tamamlandığı bir QR kod sistemi önermiştir. Baykara vd. (2017), NFC etiketlerine dayanan ve konum doğrulaması için Google Haritalar'ı kullanan mobil bir yoklama sistemi geliştirmiştir. Benyó vd. (2012) ise NFC kartlarını parmak izi doğrulamasıyla bütünleştiren karma bir çözüm sunmuştur.

Kablosuz sinyal tabanlı yaklaşımlar da bu kategoride önemli bir yer tutmaktadır. Puckdeevongs vd. (2020), BLE işaretçilerinden alınan RSSI değerlerine dayalı sınıf içi konumlandırma yapan bir Bluetooth tabanlı yoklama sistemi kurmuştur. Bhalla vd. (2013), Bluetooth MAC adresi toplamayla öğretmen cep telefonuna dayalı bir sistem önermiştir. Banepali vd. (2019), WLAN üzerinden %94'ün üzerinde doğruluk oranı elde eden bir Wi-Fi tabanlı sistem kurmuştur. Küçük vd. (2018) ise BLE ve Firebase kullanarak okul servisi takip sistemi sunmuştur.

Pasif RFID teknolojisi, düşük maliyeti ve harici güç kaynağı gerektirmemesiyle elektronik yoklama literatüründe önemli bir yer bulmaktadır (Duroc, 2022). Chiagozie ve Nwaji (2012), RFID tabanlı ilk kapsamlı sistemlerden birini geliştirmiştir. Sezdi ve Tüysüz (2018), kurumsal bilgi sistemleriyle entegre olabilen web tabanlı bir RFID yoklama sistemi sunmuştur. Aydın ve Dalkılıç (2018), öğrenci devam oranları ile akademik başarı arasındaki ilişkiyi veri madenciliğiyle ortaya koymuştur. Pala (2008) ve Özcan vd. (2018) de sırasıyla RFID tabanlı e-yoklama ve RFID+BLE entegre sistemler geliştirmiştir. Yalnızca RFID kullanan bu sistemlerin tamamında, kart devredildiğinde vekâleten yoklama sorunu varlığını sürdürmektedir.

## 2.3. Hibrit Sistemler ve Araştırma Boşluğunun Tespiti

Vekâleten yoklama sorununu gidermek amacıyla bazı çalışmalar RFID'yi biyometrik doğrulamayla birleştirmiştir. Uludağ ve Uçar (2018), Raspberry Pi tabanlı IoT altyapısında RFID ile parmak izi doğrulamasını bütünleştiren akıllı bir sınıf sistemi kurmuştur. Akbar vd. (2018), OpenCV tabanlı yüz tanımayla RFID'yi birleştiren bir sistem önermiştir. Bu hibrit yaklaşımlar vekâleten yoklama sorununu büyük ölçüde çözse de biyometrik veri depolama yükümlülüğünü ortadan kaldırmamaktadır.

Yapılan literatür taraması, mevcut çalışmaların iki temel eksiklikten en az birini barındırdığını ortaya koymaktadır: ya biyometrik veri toplamayı zorunlu kılmakta ve GDPR, KVKK ile FERPA kapsamındaki yasal riskleri beraberinde getirmekte, ya da yalnızca RFID kullanmakta ve vekâleten yoklama sorununa karşı güvence sağlayamamaktadır. Bu çalışma, söz konusu boşluğu; biyometrik veri toplamadan RFID kimlik doğrulamasını ağırlık sensörü tabanlı fiziksel varlık denetimiyle bütünleştiren özgün bir IoT mimarisiyle doldurmayı hedeflemektedir.

# 3. SİSTEM MİMARİSİ VE TASARIMI

Bu bölümde geliştirilen sistemin tasarım kararları, bileşen seçimleri ve mimari yapısı sunulmaktadır. Bölüm 5'teki fiziksel montaj ve yazılım geliştirme sürecinden farklı olarak, bu bölüm sistemin teorik çerçevesini ve özgün vekâleten yoklama tespit mekanizmasının işleyiş mantığını ele almaktadır.

## 3.1. Genel Sistem Mimarisi

Önerilen sistem, birbirini tamamlayan üç katmandan oluşan bütünleşik bir IoT mimarisi üzerine inşa edilmiştir: algılama katmanı, işleme katmanı ve yönetim katmanı. Algılama katmanında pasif RFID modülü öğrenci kimliğini tanımlarken, sandalyenin altına yerleştirilen ağırlık sensörleri fiziksel varlığı bağımsız biçimde doğrulamaktadır. İşleme katmanında Arduino UNO mikrodenetleyicisi her iki kaynaktan gelen veriyi gerçek zamanlı olarak işleyerek yoklama kararını üretmekte; yönetim katmanında ise HC-06 Bluetooth modülü aracılığıyla bilgisayara iletilen veriler Python GUI üzerinden yönetilmektedir. Şekil 1'de tüm bileşenlerin bağlantısını gösteren Fritzing devre şeması görülmektedir.

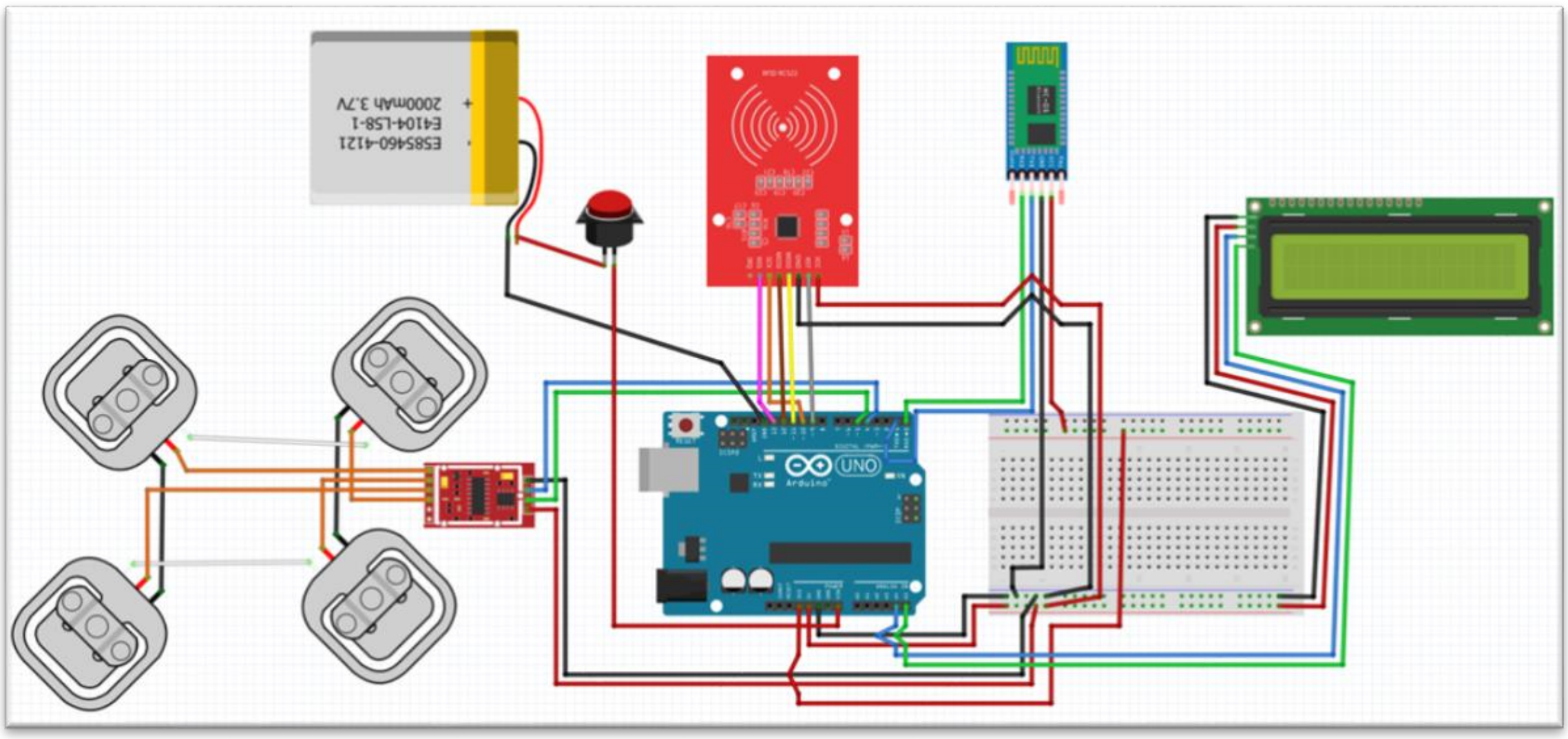

*Şekil 1. Sistemin Fritzing devre şeması (Arduino UNO, RC522, HX711, HC-06, LCD).*

Bu mimari, bilinçli olarak merkezi bir sunucu veya bulut bağlantısı gerektirmeyecek biçimde tasarlanmıştır. Bu tercih iki önemli avantaj sağlamaktadır: internet bağlantısının kesintili olduğu ortamlarda sistemin çalışmaya devam etmesi ve öğrenci verilerinin kurumun fiziksel sınırları dışına çıkmaması sayesinde veri gizliliğinin yapısal düzeyde güvence altına alınması.

## 3.2. Donanım Bileşenleri

### *3.2.1. Arduino UNO Mikrodenetleyici*

Sistemin işlem çekirdeğini ATmega328P tabanlı Arduino UNO kartı oluşturmaktadır (Badamasi, 2014; Köhli vd., 2024). 14 dijital giriş/çıkış pinine, 6 analog giriş pinine ve 32 KB program belleğine sahip bu kart (Atmel Corporation, 2015); RFID modülü, ağırlık sensörü amplifikatörü, Bluetooth modülü ve LCD ekranın eş zamanlı yönetimine olanak tanımaktadır.

### *3.2.2. RC522 RFID Modülü ve MIFARE Kartlar*

Öğrenci kimlik doğrulaması için 13,56 MHz frekansında çalışan RC522 RFID modülü tercih edilmiştir (NXP, 2016). Bu modül, ISO/IEC 14443 standardına uygun MIFARE Classic 1K kartlarla birlikte kullanılmaktadır (ISO, 2018; NXP, 2018). RC522, Arduino ile SPI protokolü üzerinden iletişim kurmaktadır ve tipik okuma mesafesi yaklaşık 3–5 cm'dir. Pasif RFID teknolojisinin kullanılması, etiketlerin harici güç gerektirmeksizin çalışmasını sağlamaktadır (Want, 2006).

### *3.2.3. HX711 ADC'li Yük Hücreleri*

Fiziksel varlık doğrulaması, sandalyenin altına yerleştirilen 50 kg kapasiteli dört adet yarı köprülü yük hücresiyle gerçekleştirilmektedir. Şekil 2'de yük hücrelerinin görünümü sunulmaktadır.

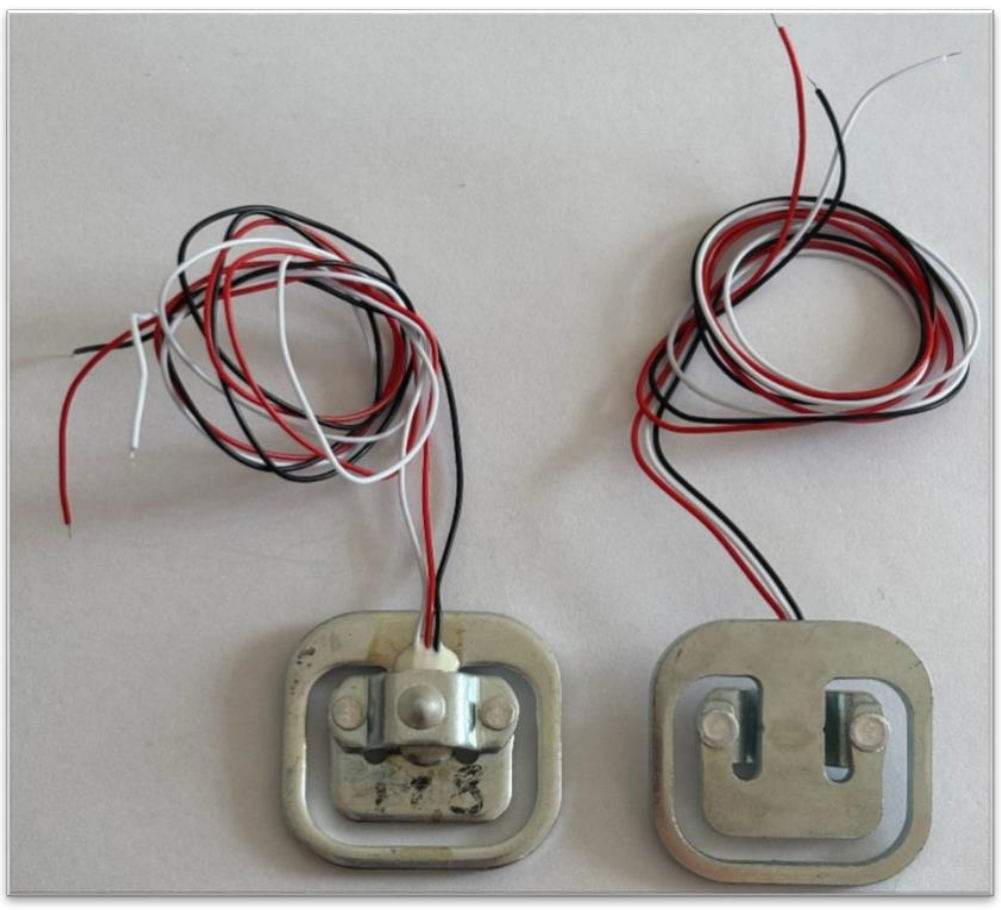

*Şekil 2. Yarı köprülü ve 50 kg kapasiteli yük hücresi.*

HX711 24-bit ADC modülü analog sinyali dijital veriye çevirerek Arduino'ya iletmektedir (SparkFun, Eylül 2016; AVIA Semiconductor). Dört sensörün Wheatstone köprüsü konfigürasyonunda bağlanması ölçüm kararlılığını artırmaktadır.

### *3.2.4. HC-06 Bluetooth Modülü*

Arduino ile bilgisayar arasındaki veri iletimi, HC-06 seri Bluetooth modülü aracılığıyla UART protokolü üzerinden kablosuz olarak gerçekleştirilmektedir. Prototipte bu modül, Arduino'dan alınan verilerin Python tabanlı grafik kullanıcı arayüzüne aktarılmasını sağlayarak sistemin işleme katmanı ile yönetim katmanı arasında kısa mesafeli ve düşük maliyetli bir haberleşme altyapısı sunmaktadır. HC-06 modülü 2,4 GHz frekans bandında çalışmakta ve açık alanda yaklaşık 10 metreye kadar etkili iletişim sağlayabilmektedir (Research Design Lab, t.y.). Şekil 3'te HC-06 Bluetooth modülünün görünümü gösterilmektedir.

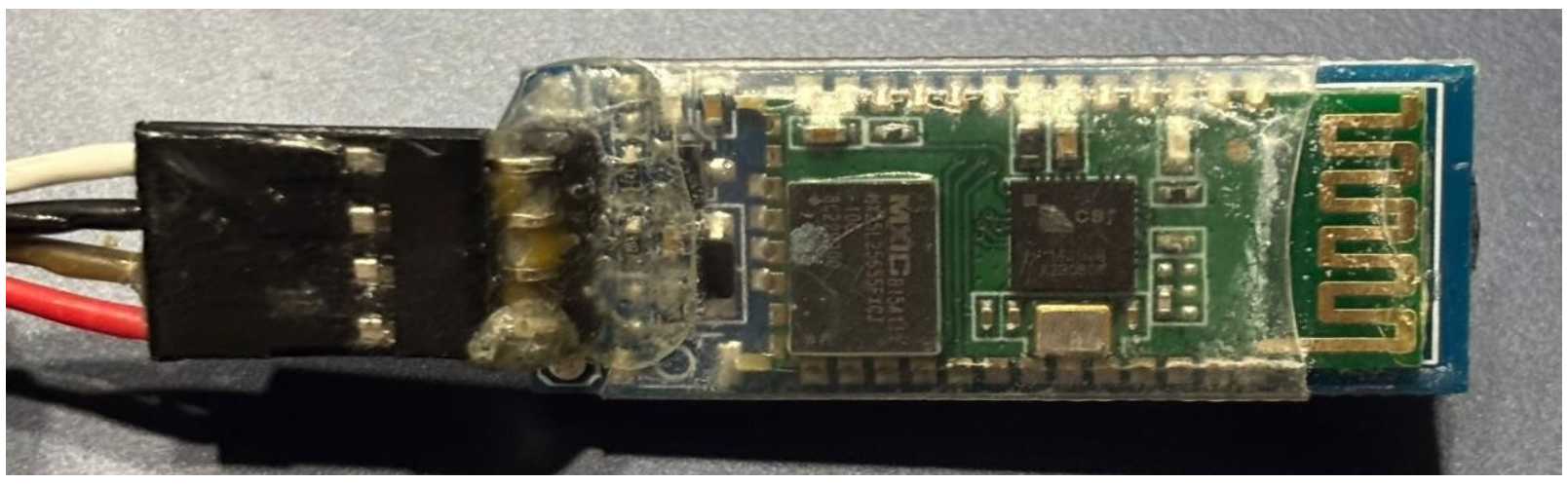

*Şekil 3. HC-06 Bluetooth modülü.*

### *3.2.5. 16×2 I2C LCD Ekran*

Öğrencilere anlık geri bildirim sağlamak amacıyla I2C arayüzlü 16×2 LCD ekran kullanılmıştır. I2C protokolü yalnızca iki iletişim hattı gerektirdiğinden Arduino'nun pin kullanımı önemli ölçüde azalmaktadır (HandsOn Tech., 2018). Şekil 4'te LCD ekranın görünümü sunulmaktadır. Ekran; başarılı okutma, tanımlanamayan kart, hatalı saat ve yoklama onayı gibi durumları anında görsel olarak bildirmektedir.

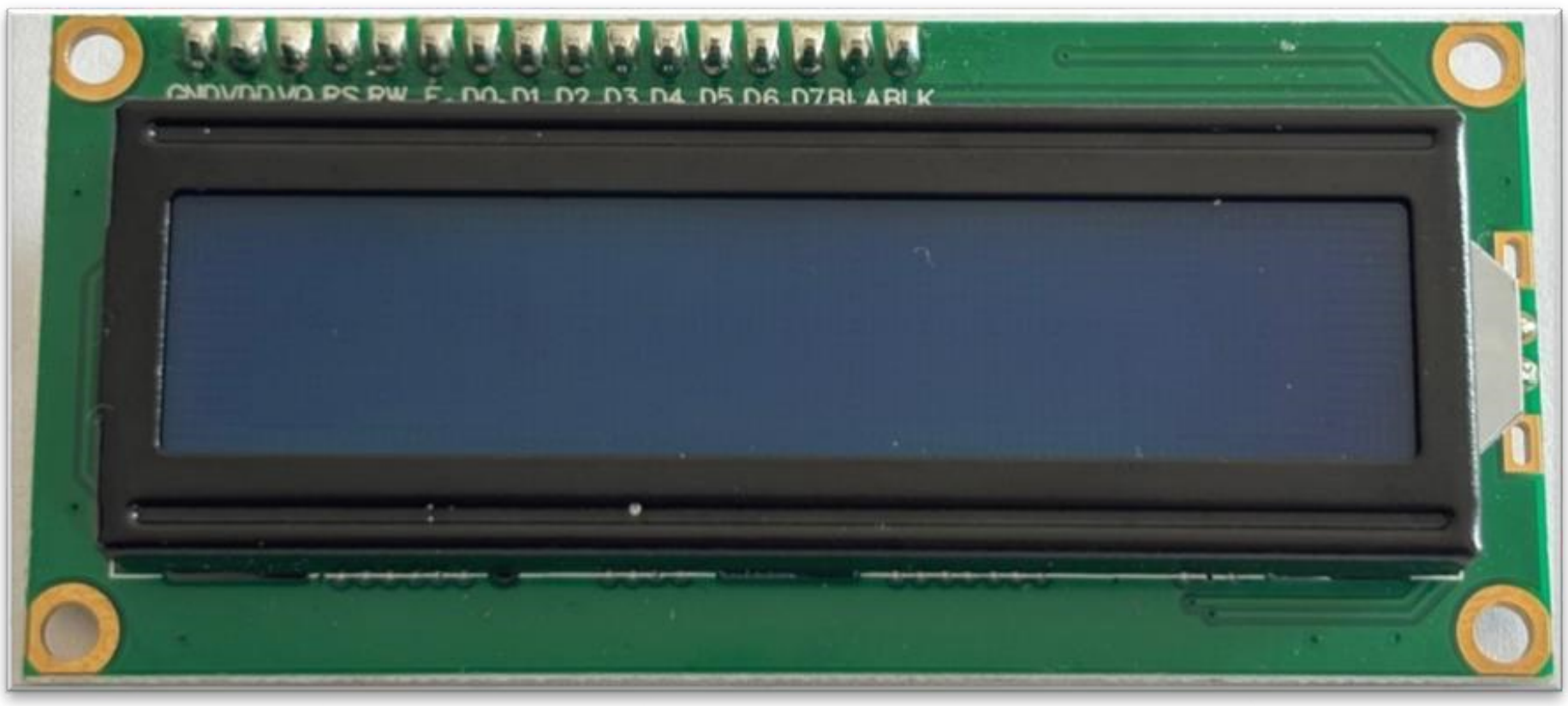

*Şekil 4. 16×2 I2C LCD ekran.*

## 3.3. Yazılım Mimarisi

### *3.3.1. Arduino Gömülü Yazılımı*

Arduino kartında çalışan gömülü yazılım Arduino IDE ortamında geliştirilmiştir (Arduino, 2024). Yazılım döngüsü üç temel görevi sırayla yerine getirmektedir: RFID okuyucudan kart kimliğini alma, ağırlık sensöründen anlık okuma değerini işleme ve bu iki veriyi birleştirerek yoklama kararını üretme. Kart okumada yeniden deneme mantığı ve sensör gürültüsü için tolerans tabanlı filtreleme algoritması uygulanmıştır.

### *3.3.2. Python GUI Uygulaması*

Bilgisayar tarafındaki yönetim arayüzü PyCharm IDE ortamında Python ile geliştirilmiştir. Uygulama; öğrenci kayıt yönetimi, ders ve saat tanımlama, Bluetooth üzerinden yoklama verilerini işleme ve kayıtları CSV formatında depolama işlevlerini yürütmektedir. CSV tabanlı depolama, kurumsal bilgi yönetim sistemleriyle entegrasyonu kolaylaştırmak amacıyla bilinçli olarak tercih edilmiştir.

## 3.4. Vekâleten Yoklama Tespit Mekanizması

Sistemin özgün katkısını oluşturan vekâleten yoklama tespit mekanizması, RFID tabanlı kimlik doğrulamasını ağırlık sensörü tabanlı fiziksel varlık denetimiyle çapraz karşılaştıran iki aşamalı bir mantığa dayanmaktadır.

Birinci aşamada sistem, RFID kartından alınan UID'yi bilgisayardaki kayıtlarla eşleştirerek kartın kayıtlı olup olmadığını, öğrencinin ilgili derse kayıtlı bulunup bulunmadığını ve okutma zamanının dersin zaman dilimiyle örtüşüp örtüşmediğini doğrulamaktadır. Eşleşme başarılı olduğunda LCD ekranda "Hoş Geldiniz, [Ad Soyad]" mesajı görüntülenmekte; doğrulama başarısız olduğunda ise "Öğrenci Bulunamadı" uyarısı verilmektedir. Şekil 5'te bu karar mantığına ait akış şeması görülmektedir.

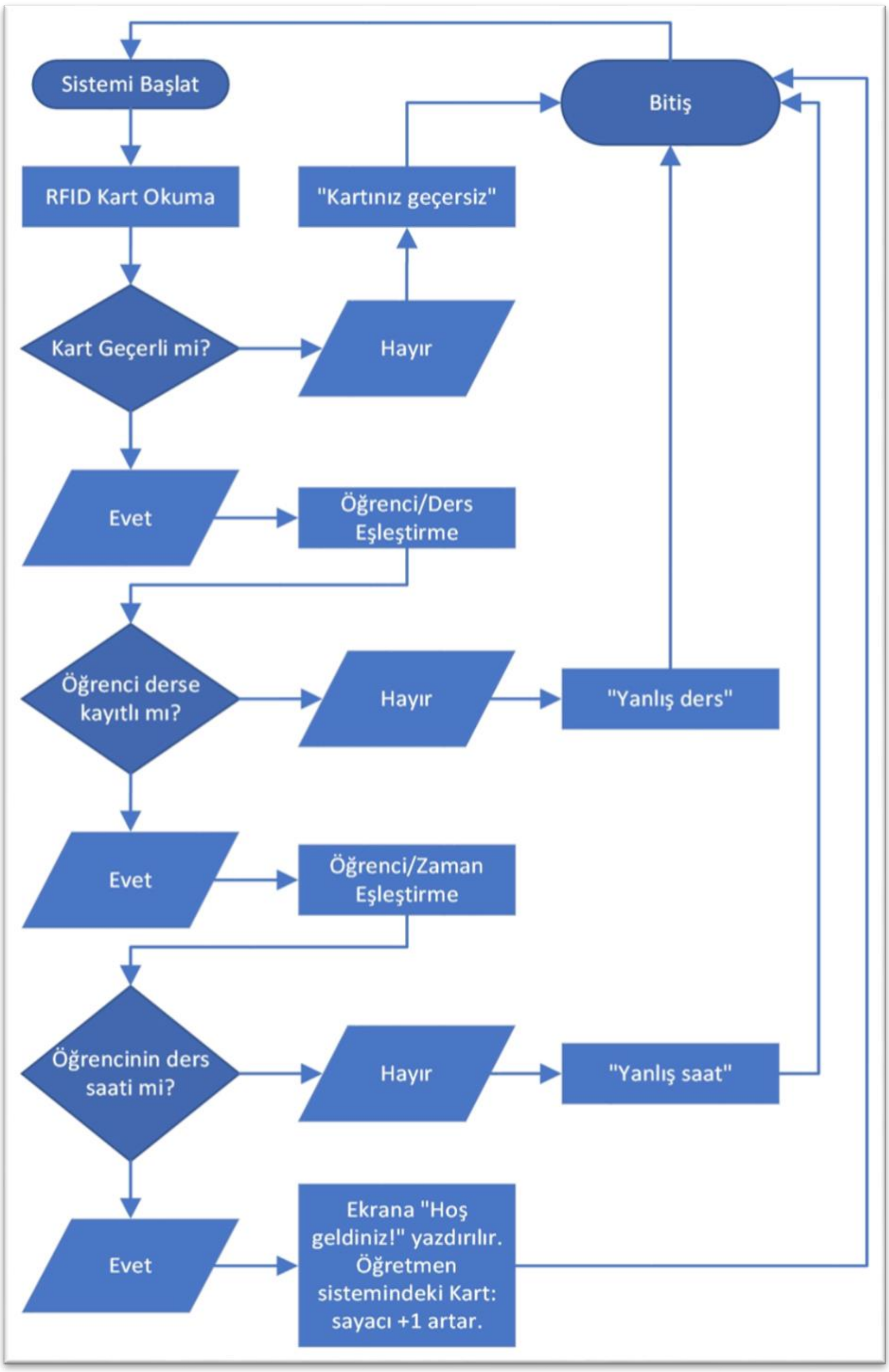

*Şekil 5. RFID kart okutma ve doğrulama akış şeması.*

İkinci aşamada ise ilgili sandalyeden alınan ağırlık değeri, istatistiksel veri setinden türetilen cinsiyet ve yaş bazlı aralıklarla karşılaştırılmaktadır. Ağırlık değerleri sisteme kaydedilmemekte, yalnızca anlık karşılaştırma amacıyla kullanılmaktadır.

Bu yaklaşım, bireysel biyometrik özellik depolamayı gerektirmemesi nedeniyle GDPR, KVKK ve FERPA bağlamındaki veri koruma yükümlülüklerini önemli ölçüde azaltmaktadır. Şekil 6'da ağırlık sensörü doğrulama sürecinin akış şeması verilmektedir.

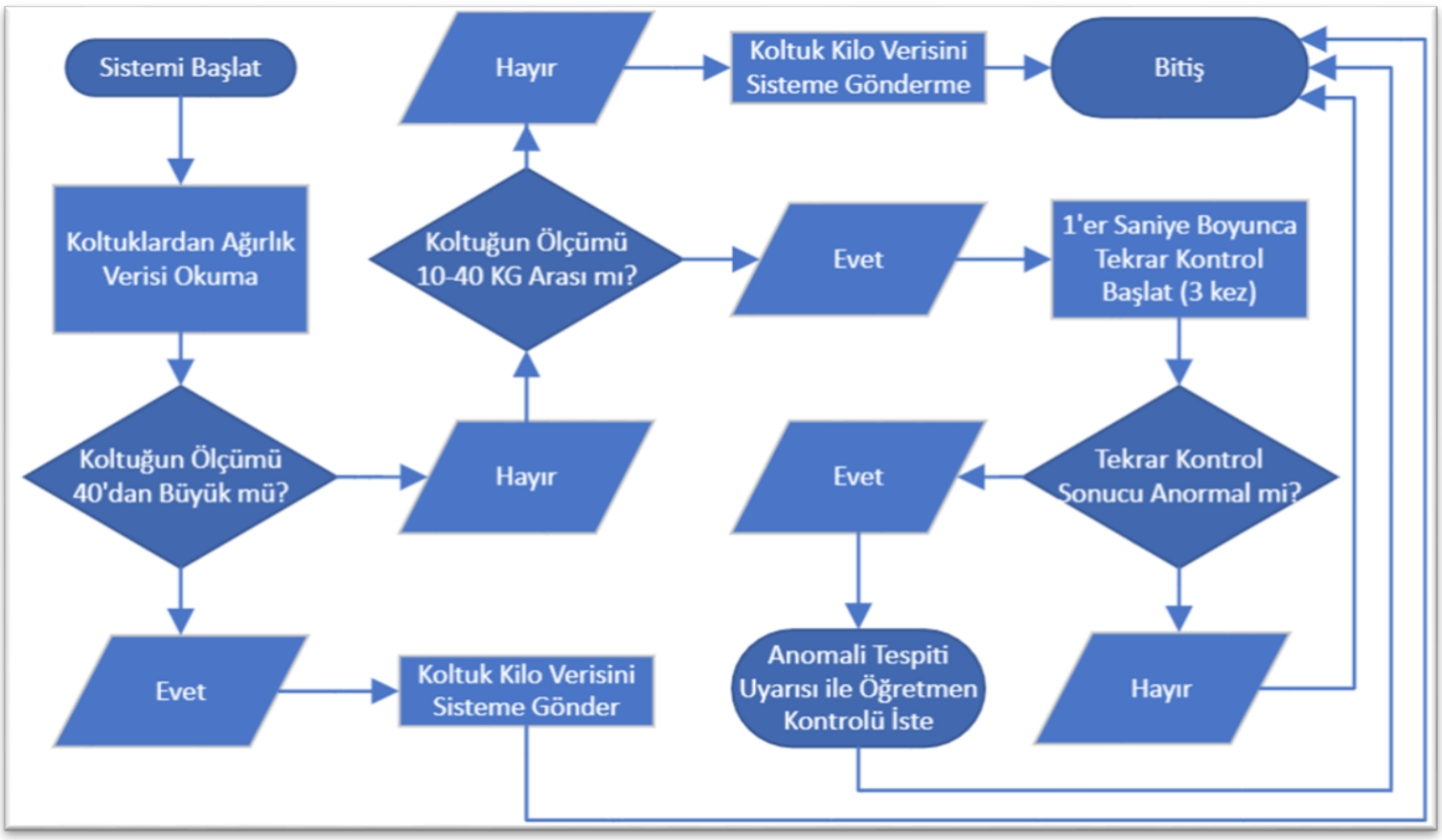

*Şekil 6. Ağırlık sensörü doğrulama akış şeması.*

# 4. AĞIRLIK SENSÖRÜ DOĞRULAMASINA YÖNELİK VERİ ANALİZİ

Bu bölümde, Bölüm 3'te tasarlanan ağırlık sensörü doğrulama mekanizmasının karar eşiklerini belirlemek amacıyla yürütülen istatistiksel analiz sunulmaktadır. Analiz, Bölüm 5'teki fiziksel uygulamadan ve Bölüm 6'daki test sürecinden bağımsız olarak, sistemin teorik doğrulama altyapısını oluşturmaktadır.

## 4.1. Veri Toplama

Ağırlık sensörü tabanlı varlık doğrulama mekanizmasının güvenilir işleyebilmesi için hedef kullanıcı kitlesini temsil eden istatistiksel bir ağırlık referansına ihtiyaç duyulmaktadır. Ulusal kaynaklarda 18–22 yaş grubuna özgü yeterli veri bulunamamış; bunun üzerine uluslararası açık veri platformu Kaggle'dan üç ayrı veri seti derlenmiştir.

Gym Members Exercise Dataset (Kaggle, 2024) 973 kayıt; Obesity Classification Dataset (Kaggle, 2023) 109 kayıt; Medical Cost Personal Datasets (Kaggle, 2018) ise 1.338 kayıt içermektedir. Son veri setinde ağırlık değerleri doğrudan verilmediğinden, erkekler için 170 cm ve kadınlar için 160 cm sabit boy varsayımı altında BMI = Ağırlık / Boy$^2$ formülünden hesaplanmıştır.

## 4.2. Veri Ön İşleme

Her üç veri seti üzerinde Python programlama dili ile Pandas ve Matplotlib kütüphaneleri kullanılarak sistematik bir veri temizliği yürütülmüştür. Yalnızca yaş, cinsiyet ve ağırlık sütunları korunmuş; 18–22 yaş dışındaki bireyler ve 40 kg altındaki değerler çıkarılmıştır. Üç filtrelenmiş veri seti birleştirilmiş ve toplam 350 bireylik nihai örneklem elde edilmiştir.

## 4.3. İstatistiksel Bulgular

***Cinsiyet Dağılımı***

Nihai örneklemin %51,4'ünü (180 birey) erkekler, %48,6'sını (170 birey) kadınlar oluşturmaktadır (Şekil 7). Her iki cinsiyetin neredeyse eşit oranlarda temsil edilmesi, analiz bulgularının dengeli biçimde yorumlanmasına olanak tanımaktadır.

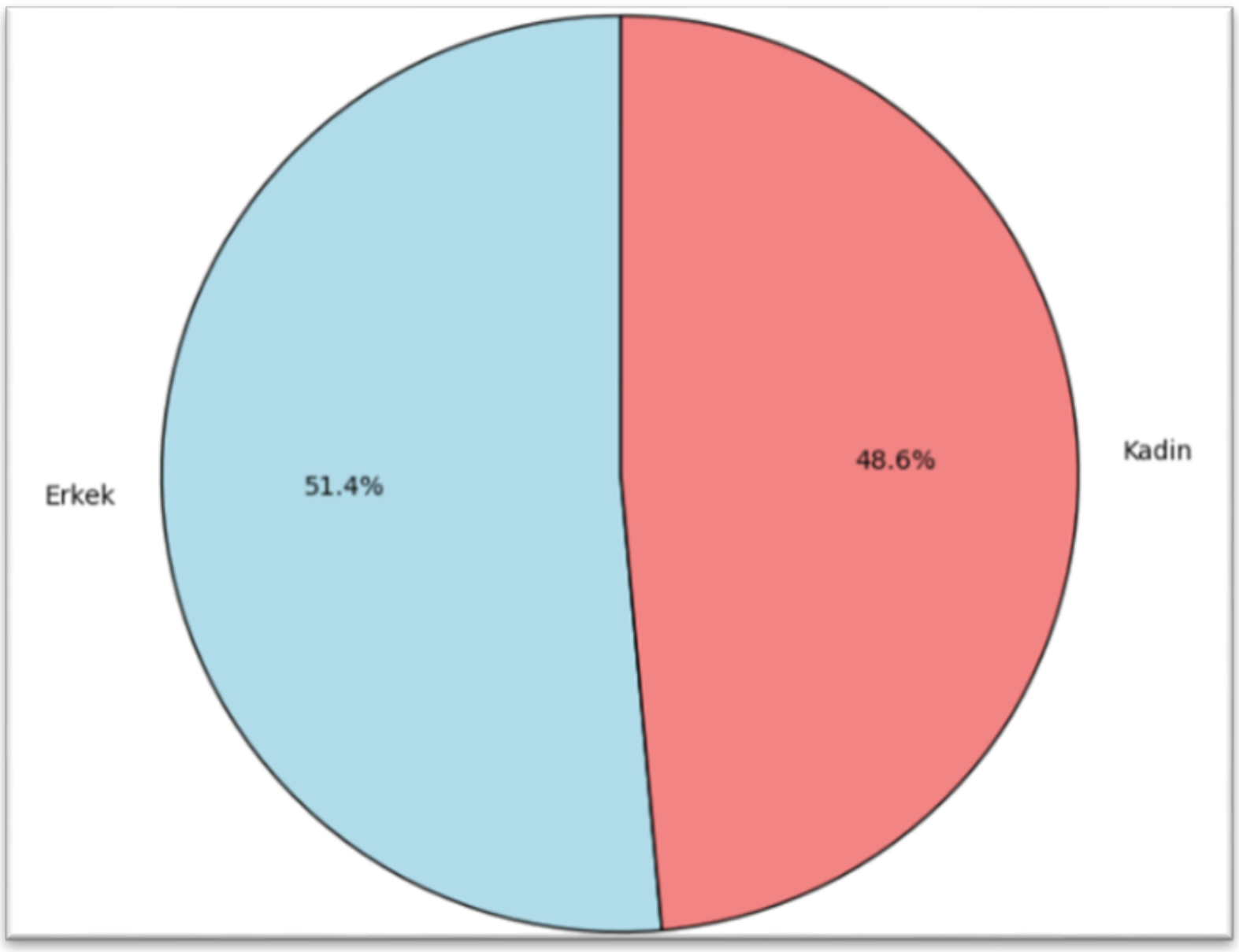

*Şekil 7. Örneklemin cinsiyet dağılımı (n=350; Erkek: %51,4, Kadın: %48,6).*

***Cinsiyete ve Yaşa Göre Ortalama Ağırlık Değerleri***

Tüm yaş gruplarında erkeklerin ortalama ağırlığı kadınlardan belirgin biçimde yüksektir (Şekil 8). Erkeklerde yaşa göre ortalama değerler şöyledir: 18 yaşında 85,41 kg, 19 yaşında 82,76 kg, 20 yaşında 94,59 kg, 21 yaşında 85,44 kg ve 22 yaşında 92,52 kg. Kadınlarda ise bu değerler sırasıyla 76,96 kg, 69,37 kg, 70,20 kg, 66,86 kg ve 65,11 kg olarak gerçekleşmiştir.

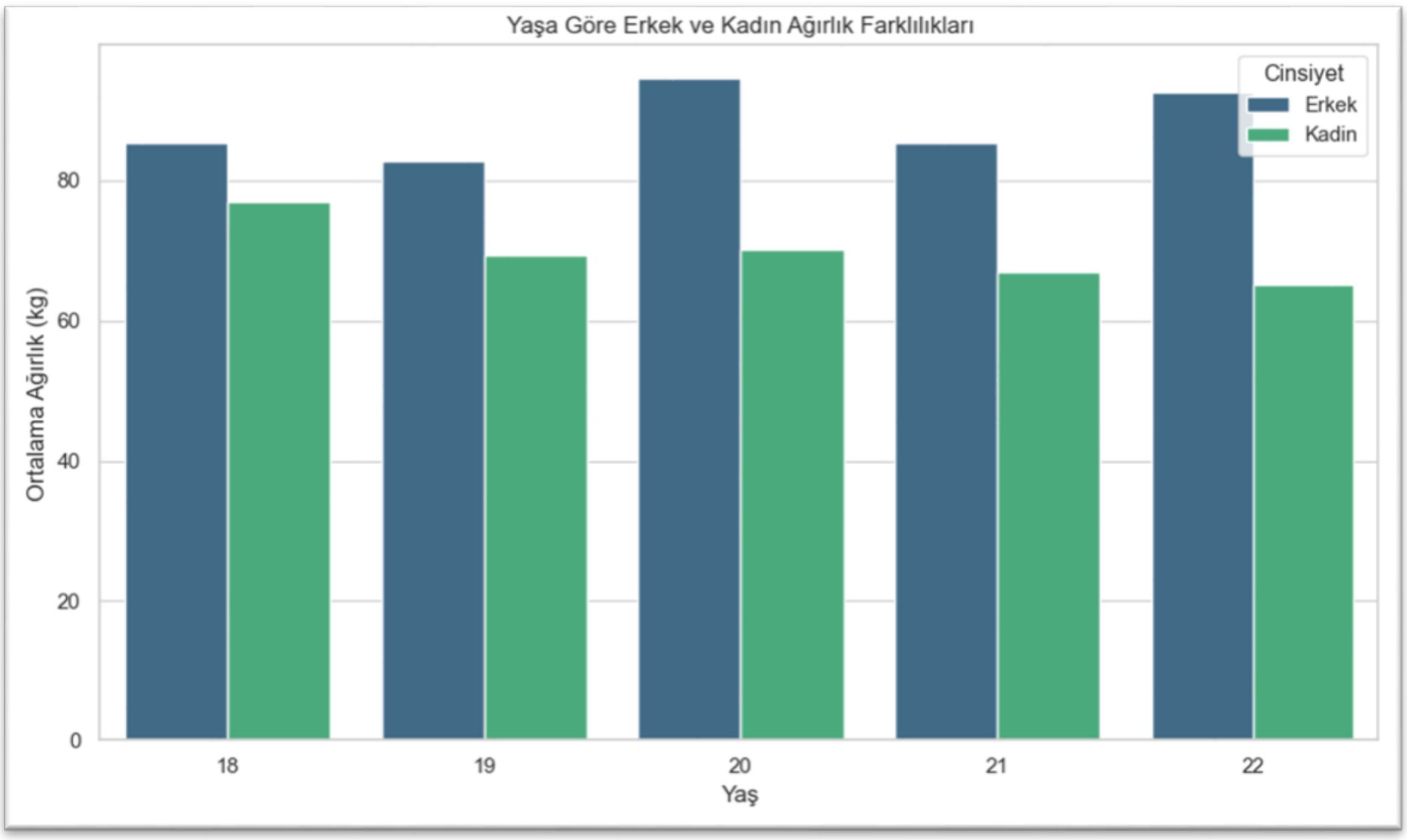

*Şekil 8. Yaşa ve cinsiyete göre ağırlık dağılımları (18–22 yaş grubu, n=350).*

Şekil 9'da yaş gruplarına göre ortalama ağırlık değerleri görülmektedir.

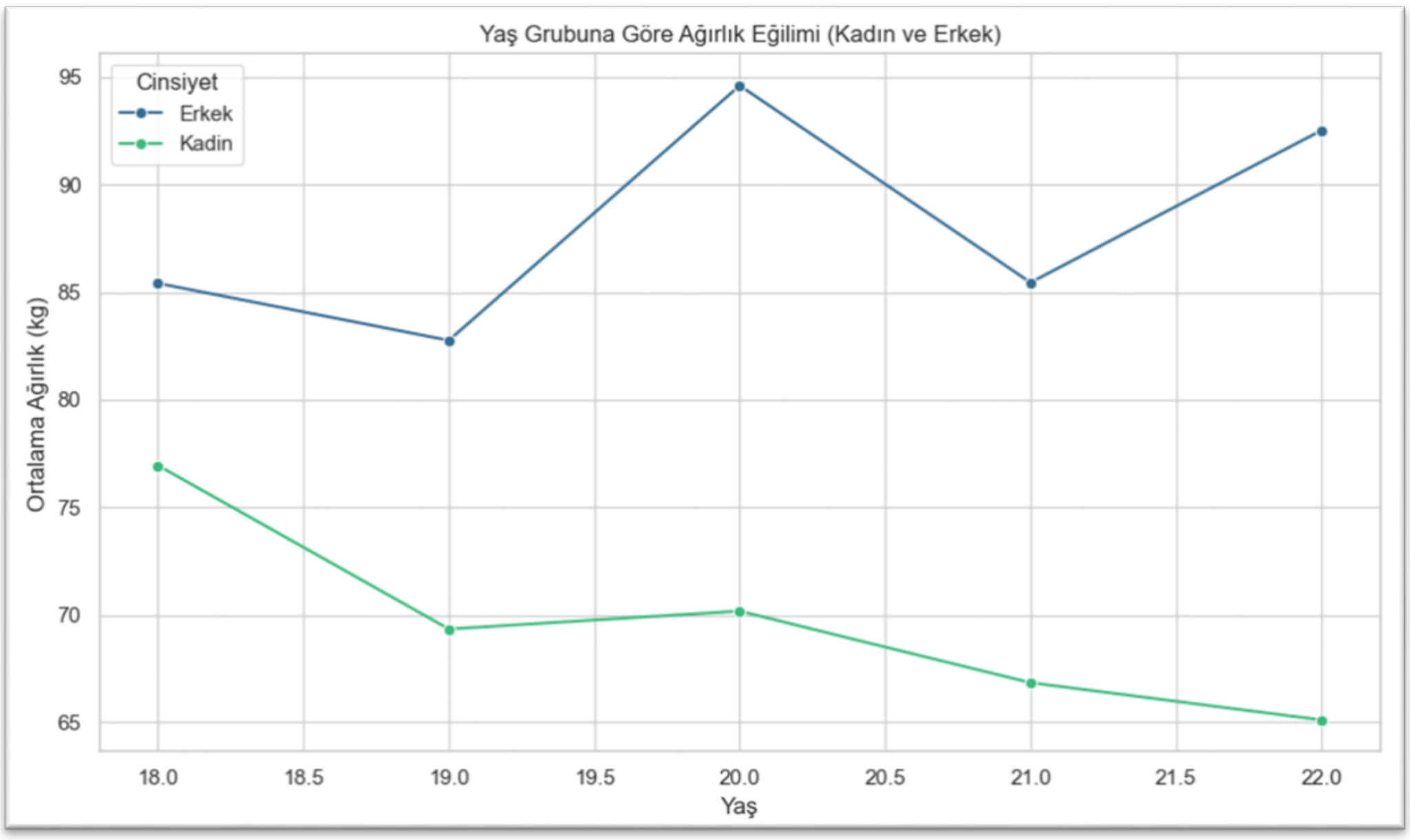

*Şekil 9. Yaş gruplarına göre ortalama ağırlık değerleri (18–22 yaş grubu, n=350).*

### *Cinsiyetler Arası Ağırlık Farkı*

18 yaş grubunda erkeklerin ortalama ağırlığı kadınlara kıyasla yaklaşık %11 daha yüksekken, bu fark 20 yaş grubunda %35'e çıkmaktadır. Bu bulgu, ağırlık eşiklerinin cinsiyete göre ayrı belirlenmesinin gerekliliğini doğrulamakta ve sistemde cinsiyet bazlı eşik uygulamasını zorunlu kılmaktadır.

### 4.4. Doğrulama Eşiklerinin Sisteme Entegrasyonu

Bölüm 4 analizinden elde edilen cinsiyet ve yaş bazlı ortalama ağırlık değerleri, sistemde "kilo havuzu" mekanizmasının istatistiksel temelini oluşturmaktadır. Bir öğrenci RFID kartını başarıyla okuttuğunda, ilgili yaş ve cinsiyet grubuna ait ortalama ağırlık değeri sisteme referans olarak eklenmektedir. Bu sayede sınıftaki kayıtlı öğrenciler için beklenen toplam ağırlık hesaplanabilmektedir. Sandalye sensörlerinden ölçülen gerçek toplam ağırlık ise bu beklenen değerle karşılaştırılmakta, belirlenen tolerans sınırının dışında kalan farklar öğretmene anomali uyarısı olarak bildirilmektedir.

Bireysel sandalye düzeyinde, 10 ile 40 kg arasındaki geçici ölçümler oturma ve kalkma geçişleri olarak değerlendirilmekte ve 1 saniyelik aralıklarla üç kez yeniden kontrol edilmektedir. 40 kg üzerindeki ölçümler ise aktif oturma durumu olarak kabul edilmektedir. Böylece hem bireysel sandalye düzeyindeki eşik kontrolü hem de sınıf toplamına dayalı istatistiksel karşılaştırma birlikte kullanılarak sistemin sensör gürültüsüne karşı dayanıklılığı artırılmaktadır.

Bu çalışmada tanımlanan kilo havuzu mekanizması, çok sandalyeli tam sınıf mimarisi için öngörülen doğrulama mantığını temsil etmektedir. Bununla birlikte, bu makalede sunulan prototip kavram kanıtı düzeyinde tek sandalye ile sınırlandırılmıştır. Bu nedenle sınıf geneli toplam ağırlık karşılaştırması henüz tam ölçekli donanım üzerinde uygulanmamıştır.

Bu yaklaşımın temel gizlilik avantajı şudur: sisteme kaydedilen veriler yalnızca öğrencinin yaşı ve cinsiyetidir. Bu bilgiler, kurumsal kayıt sistemlerinde zaten bulunan ve özel nitelikli kişisel veri kapsamında değerlendirilmeyen standart demografik verilerdir. Anlık ağırlık ölçümü ise sisteme kaydedilmemekte, yalnızca bir kerelik karşılaştırma amacıyla kullanılmaktadır.

## 5. UYGULAMA VE PROTOTİP

Bu bölümde, Bölüm 3'te tasarlanan sistemin fiziksel olarak hayata geçirilme süreci ayrıntılandırılmaktadır. Bölüm 3'teki tasarım kararlarından farklı olarak, bu bölüm donanımın gerçek montajını, bileşenler arasındaki kablolama bağlantılarını, yazılım geliştirme adımlarını ve veri akış yapısını ele almaktadır.

## 5.1. Donanım Kurulumu

Prototip, tek bir Arduino UNO kartı etrafında şekillenen modüler bir donanım mimarisine sahiptir. Dört adet yarı köprülü yük hücresi lehim yöntemiyle HX711 ADC modülüne bağlanmış; oluşturulan bütün Arduino'nun dijital pinlerine entegre edilmiştir. Şekil 10'da lehim işlemi tamamlanmış HX711 modülü ve bağlı sensörler görülmektedir.

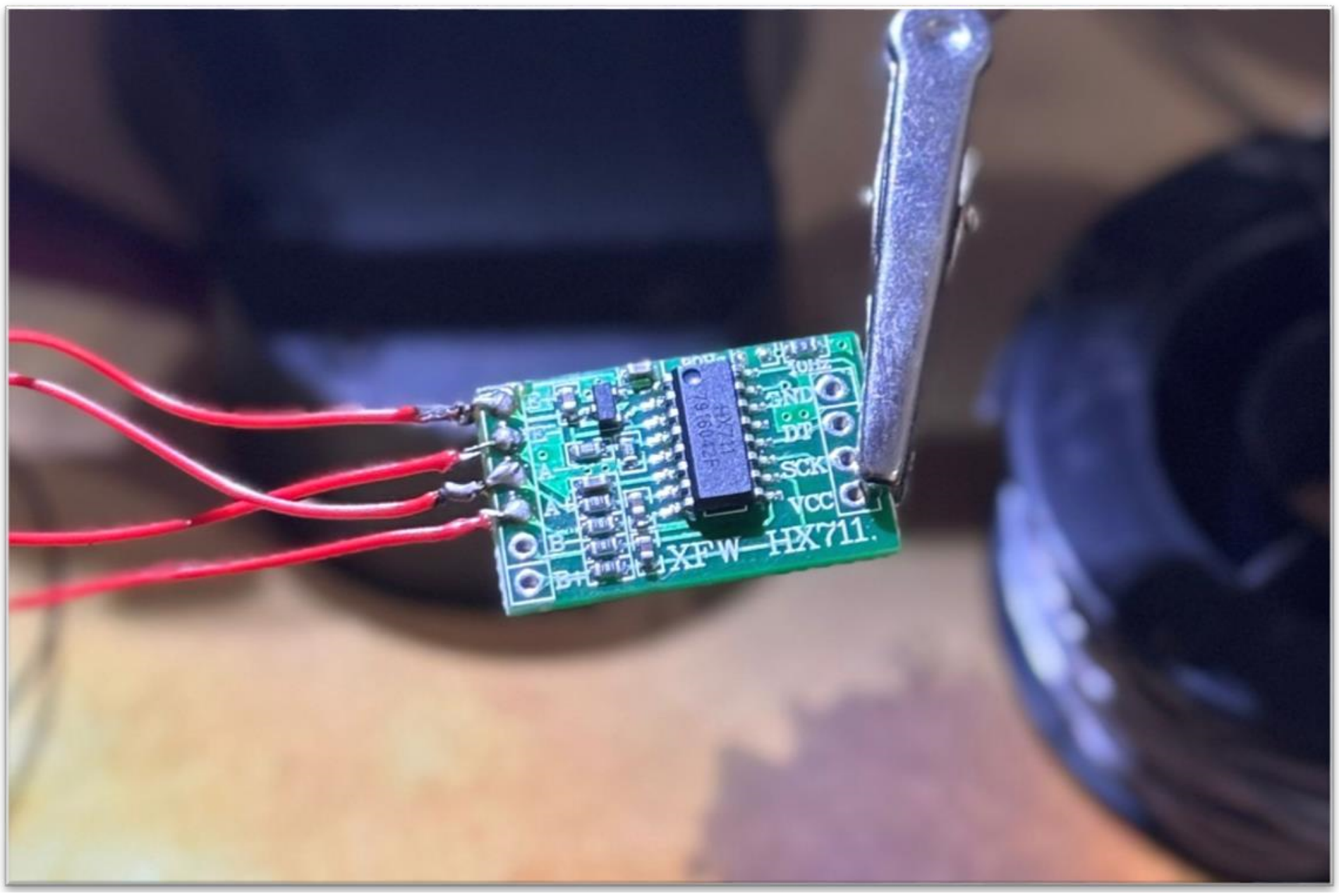

*Şekil 10. Dört ağırlık sensörü lehimlenmiş HX711 ADC modülü.*

Sensörler, öğrenci oturma yükünü doğrudan karşılayacak biçimde sandalye oturma tahtasının alt köşelerine yerleştirilmiştir.

RC522 RFID modülünün IRQ dışındaki tüm pinlerine renk kodlu atlama kabloları takılmış; SDA pini Arduino UNO'nun D10 pinine, SCK pini D13 pinine, MOSI pini D11 pinine ve MISO pini D12 pinine bağlanmıştır. Şekil 11'de bağlantıları tamamlanmış RC522 modülü görülmektedir.

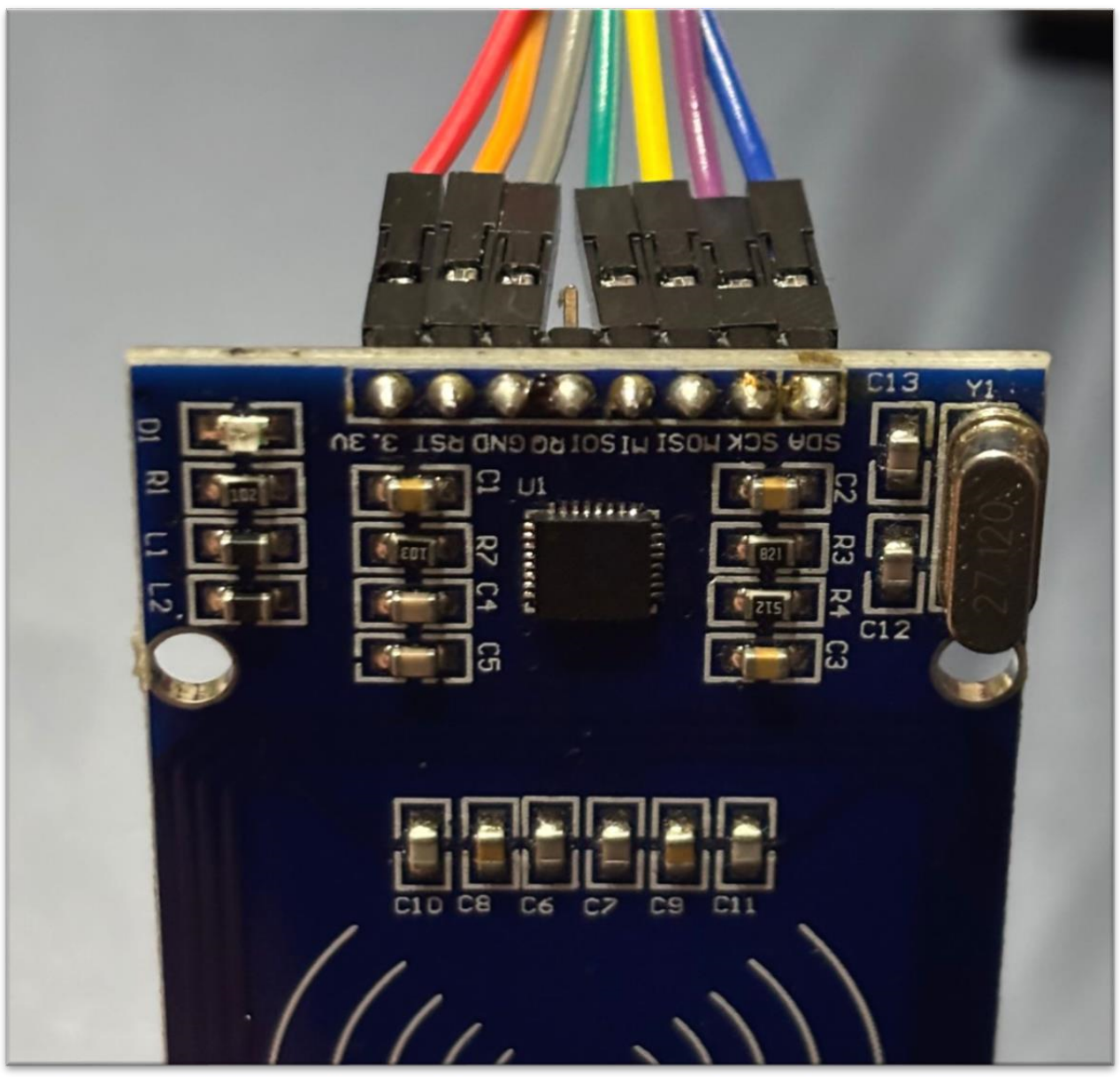

*Şekil 11. Renk kodlu kablolarla bağlanmış RC522 RFID modülü.*

HC-06 Bluetooth modülünün VCC pini Arduino'nun 3,3V pinine, GND pini GND pinine, TX pini RX (D0) pinine ve RX pini TX (D1) pinine bağlanmıştır. I2C modüllü LCD ekran yalnızca dört iletişim hattıyla Arduino'ya bağlanmıştır. Tüm bileşenler Arduino'nun 5V çıkışıyla beslendiğinden harici güç kaynağı gereksizdir. Şekil 12'de tamamlanmış prototip donanımının genel görünümü sunulmaktadır.

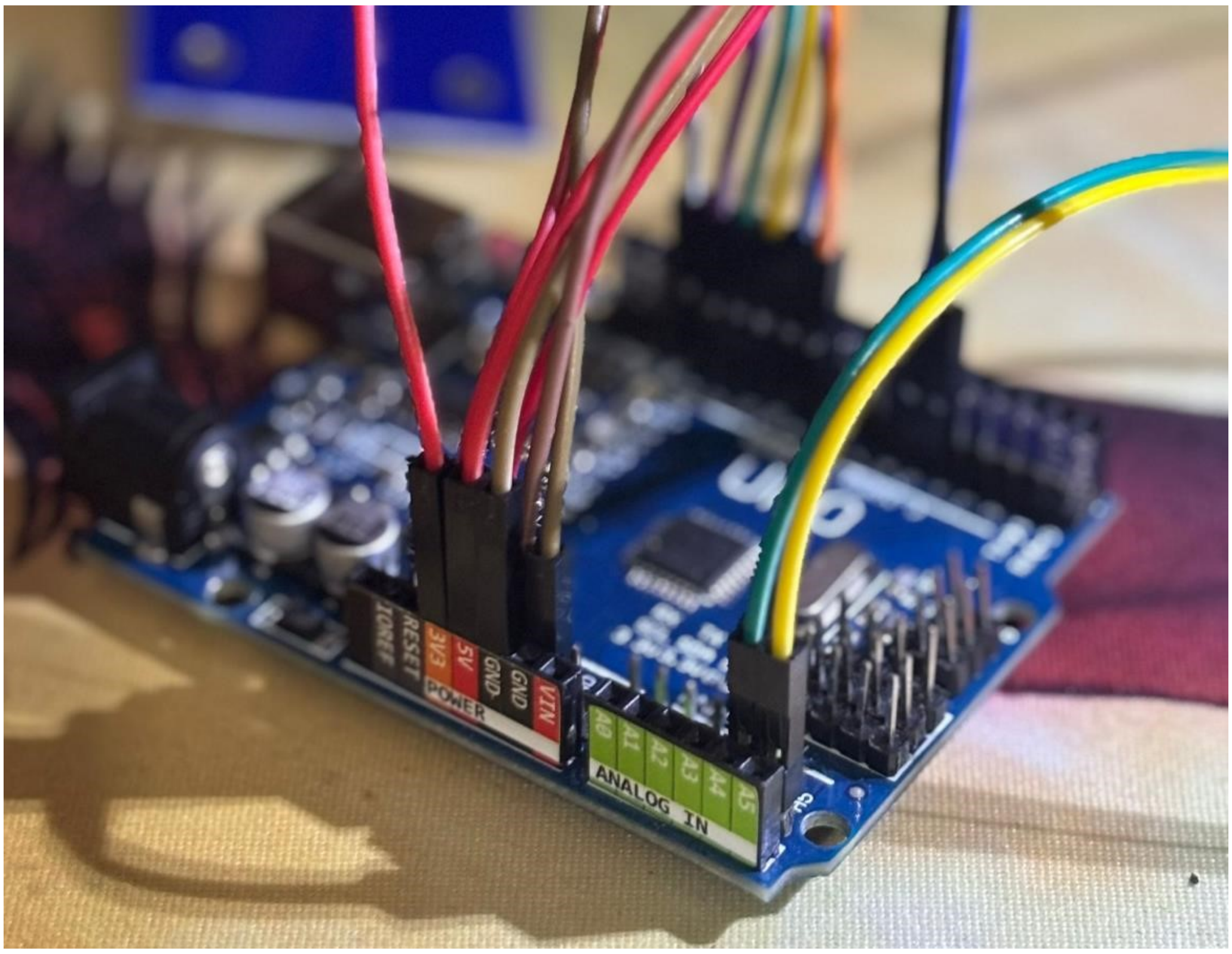

*Şekil 12. Gerçekleştirilmiş prototip donanımının genel görünümü.*

## 5.2. Yazılım Geliştirme ve Entegrasyon

***Arduino Gömülü Yazılımı***

Arduino gömülü yazılımı üç temel modülden oluşmaktadır: RFID okuma (yeniden deneme mantığı dahil), ağırlık ölçümü (ADC değerlerinin kilogram birimine dönüştürülmesi) ve Bluetooth iletişimi. Ana döngü yaklaşık 10 Hz frekansta çalışmakta; bu değer yanıt hızı ile kararlılık arasında uygun bir denge sağlamaktadır.

***Python GUI Uygulaması***

Masaüstü yönetim arayüzü dört işlevsel modül içermektedir: öğrenci yönetimi, ders yönetimi, yoklama ve raporlama. Her öğrenci kaydı; ad soyad, RFID kart kimliği (UID), yaş ve cinsiyet bilgilerini kapsamaktadır. Yoklama modülü Bluetooth üzerinden gelen verileri gerçek zamanlı işlemekte ve doğrulanmış girişleri CSV dosyasına yazmaktadır. Şekil 13'te GUI ana ekranı görülmektedir.

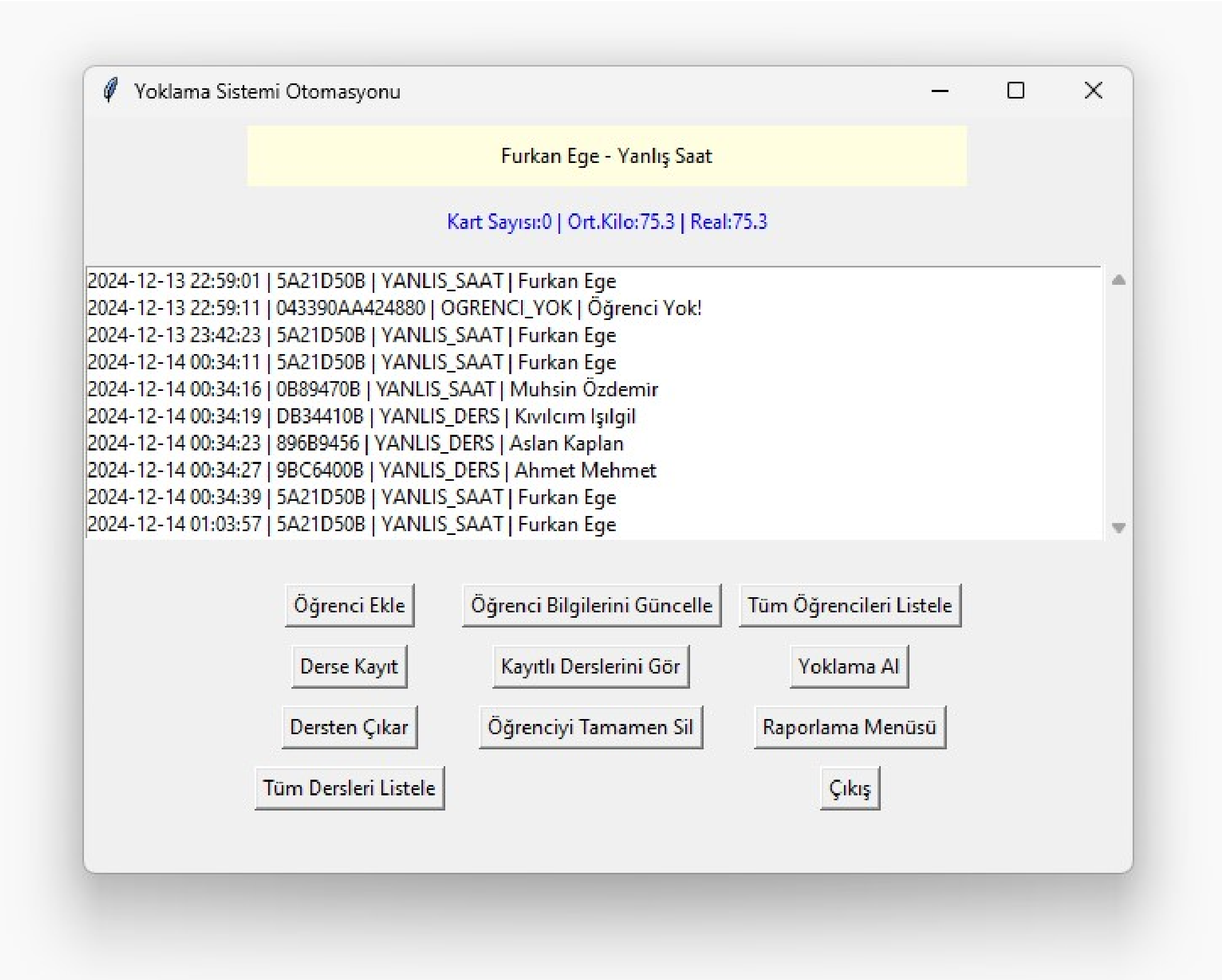

*Şekil 13. Python GUI ana ekranı: yoklama kayıtları, öğrenci ve ders yönetim işlevleri.*

***Prototip Aşamasında Yaş Bilgisi***

Prototip aşamasında öğrenci yaş bilgisi, sistemin işlev gösterdiğinin doğrulanması amacıyla 18–22 aralığından simüle edilmiştir; mevcut GUI formunda yaş alanı ayrıca yer almamaktadır. Gerçek kurumsal uygulamada bu değer, üniversite bilgi sistemlerinden otomatik olarak alınacak ve Bölüm 4'teki istatistiksel referans tablosuyla doğrudan eşleştirilecektir.

Bu tasarım kararı, sistemin kavram kanıtı (proof-of-concept) düzeyindeki mevcut kapsamını sınırlamakta olmakla birlikte, yaş bazlı doğrulama mantığının temel mekanizmasını etkilememektedir.

***Veri Depolama Yapısı***

Sistem, kurumsal veri tabanı yazılımı gerektirmeyen CSV tabanlı bir depolama mimarisini benimsemektedir. Öğrenci kayıtları, ders bilgileri ve zaman damgalı yoklama girişleri ayrı dosyalarda tutulmakta; bu yapı mevcut üniversite bilgi sistemleriyle içe/dışa aktarma uyumluluğunu kolaylaştırmaktadır.

## 5.3. Sistem İş Akışı

Yoklama alma döngüsü şu sırayla işlemektedir: öğretim elemanı GUI üzerinden ilgili dersi seçerek yoklama oturumunu başlatır. Sistem, yalnızca geçerli zaman dilimiyle eşleşen ders için yoklama kaydı oluşturulmasına izin vermektedir.

Bir öğrenci RFID kartını okuttuğunda, sistem önce Bölüm 3.4’te tanımlanan koşulları sırayla denetlemektedir. Kartın sisteme kayıtlı olması, öğrencinin ilgili derse kayıtlı bulunması ve okutma zamanının ders saatine uygun olması durumunda ağırlık doğrulama aşamasına geçilmektedir. Bu aşamada ölçülen değer, Bölüm 4’te açıklanan yaş ve cinsiyet temelli istatistiksel referans mantığı çerçevesinde değerlendirilmektedir.

Tolerans aralığı içinde kalan ölçümlerde yoklama onaylanmakta ve işlem zaman damgasıyla birlikte CSV dosyasına kaydedilmektedir. Böylece sistem, kimlik doğrulama ile fiziksel varlık denetimini ardışık biçimde birleştirerek yoklama kararını oluşturmaktadır. Tüm iş akışı saniyeler içinde tamamlanmakta ve ders süresini en düşük düzeyde etkilemektedir.

# 6. TEST SÜRECİ VE BULGULAR

Bu bölümde, Bölüm 5'te gerçekleştirilen prototip üzerinde yürütülen test süreci ve elde edilen bulgular aktarılmaktadır. Bölüm 5'in montaj ve geliştirme sürecini ele almasından farklı olarak bu bölüm, sistemin nasıl çalıştığını ve hangi sorunların nasıl giderildiğini belgeleyen test sonuçlarına odaklanmaktadır.

## 6.1. Test Yaklaşımı

Sistemin işlevsel doğruluğunu ve bütünleşik performansını değerlendirmek amacıyla iki aşamalı bir test stratejisi benimsenmiştir. Birinci aşamada her modül bağımsız olarak test edilmiş, ikinci aşamada ise modüller birlikte çalıştırılarak uçtan uca sistem davranışı gerçek sınıf koşullarına yakın bir ortamda sınanmıştır. Tablo 1'de test edilen senaryolar, beklenen davranışlar ve gözlenen sonuçlar özetlenmektedir. Bu testler, prototipin işlevsel olarak beklenen davranışı sergileyip sergilemediğini değerlendirmeye yönelik nitel doğrulama testleri olarak yürütülmüştür. Sistematik nicel başarı oranları ve hata metrikleri ise bu çalışmanın kapsamında raporlanmamıştır.

*Tablo 1. Test senaryoları, beklenen davranışlar ve gözlenen sonuçlar.*

| Test Senaryosu | Beklenen Davranış | Gözlenen Sonuç |
|---|---|---|
| Kayıtlı kart + doğru ders saati | LCD ekranında "Hoş Geldiniz" mesajının gösterilmesi ve yoklama kaydının oluşturulması | Beklenen davranış gözlendi |
| Kayıtsız kart | LCD ekranında "Öğrenci Bulunamadı" uyarısının gösterilmesi ve kayıt oluşturulmaması | Beklenen davranış gözlendi |
| Kayıtlı kart + yanlış ders saati | LCD ekranında "Yanlış Saat" uyarısının gösterilmesi ve kayıt oluşturulmaması | Beklenen davranış gözlendi |
| Kayıtlı kart + kayıtsız ders | LCD ekranında "Yanlış Ders" uyarısının gösterilmesi ve kayıt oluşturulmaması | Beklenen davranış gözlendi |
| Aynı kartın tekrar okutulması | Kimlik doğrulamanın tutarlı biçimde sürdürülmesi ve bellek kararlılığının korunması | Beklenen davranış gözlendi |
| Farklı referans ağırlıklar | Kalibrasyon sonrasında ölçümlerin kilogram değerleriyle tutarlı biçimde eşleşmesi | Beklenen davranış gözlendi |
| Arduino IDE monitörü açıkken | GUI'nin COM portuna erişememesi durumunun tespit edilmesi ve uygun kullanım protokolünün belirlenmesi | Protokol belirlendi |
| Uzun süreli çalıştırma | Bellek, port ve dosya işlemlerinin kararlı biçimde sürdürülmesi | Beklenen davranış gözlendi |

## 6.2. Modül Düzeyinde Testler

***RFID Okuma Modülü***

RC522 modülü hem kayıtlı hem de kayıtsız kartlarla test edilmiştir. Kayıtlı kartlarda sistem tutarlı biçimde kart kimliğini doğru tanımlamış; Şekil 14'te görülen başlangıç ekranının ardından Şekil 15'teki onay mesajı LCD'de görüntülenmiş ve Bluetooth üzerinden veri iletilmiştir.

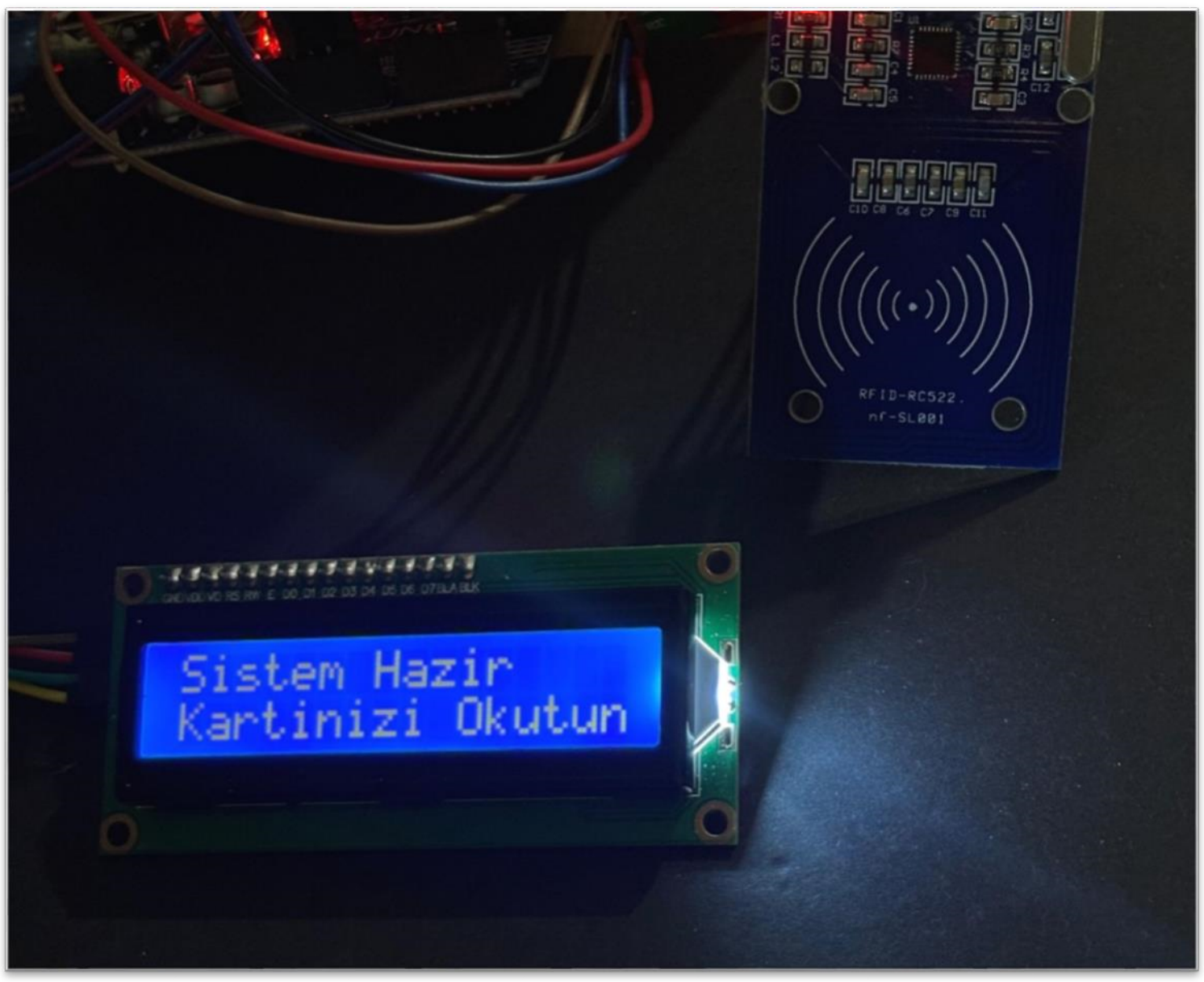

*Şekil 14. Sistem başlangıcındaki LCD görüntüsü: "Sistem Hazır, Kartınızı Okutun".*

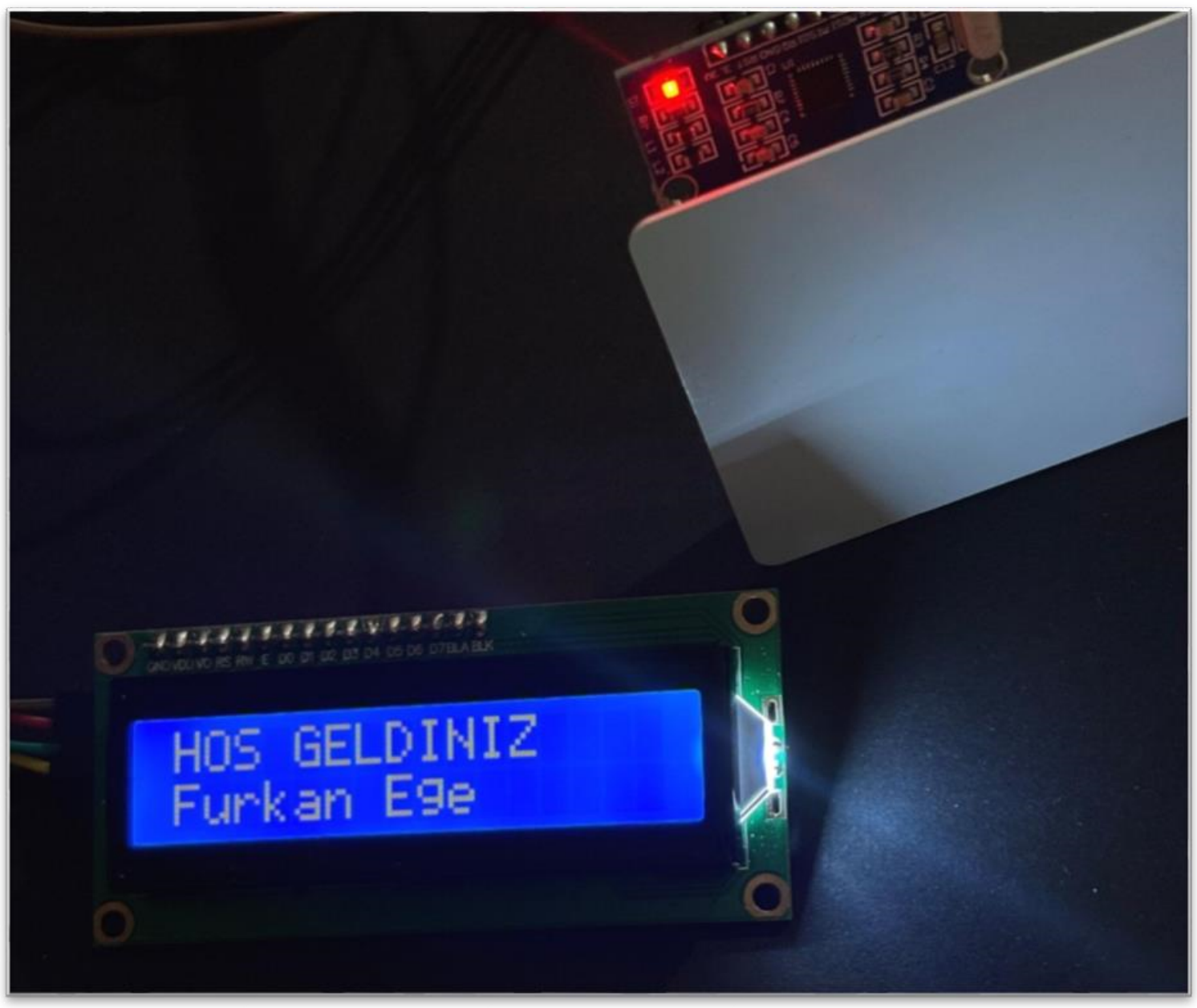

*Şekil 15. Kayıtlı kart okutulması durumunda LCD görüntüsü: "Hoş Geldiniz".*

Kayıtsız kartlarda ise Şekil 16'da görülen "Öğrenci Bulunamadı" uyarısı ekrana yansımış ve yoklama kaydı oluşturulmamıştır. Kart okumanın zaman zaman ilk denemede tamamlanamadığı gözlemlenmiş; yazılıma eklenen yeniden deneme mantığıyla bu sorun giderilmiştir.

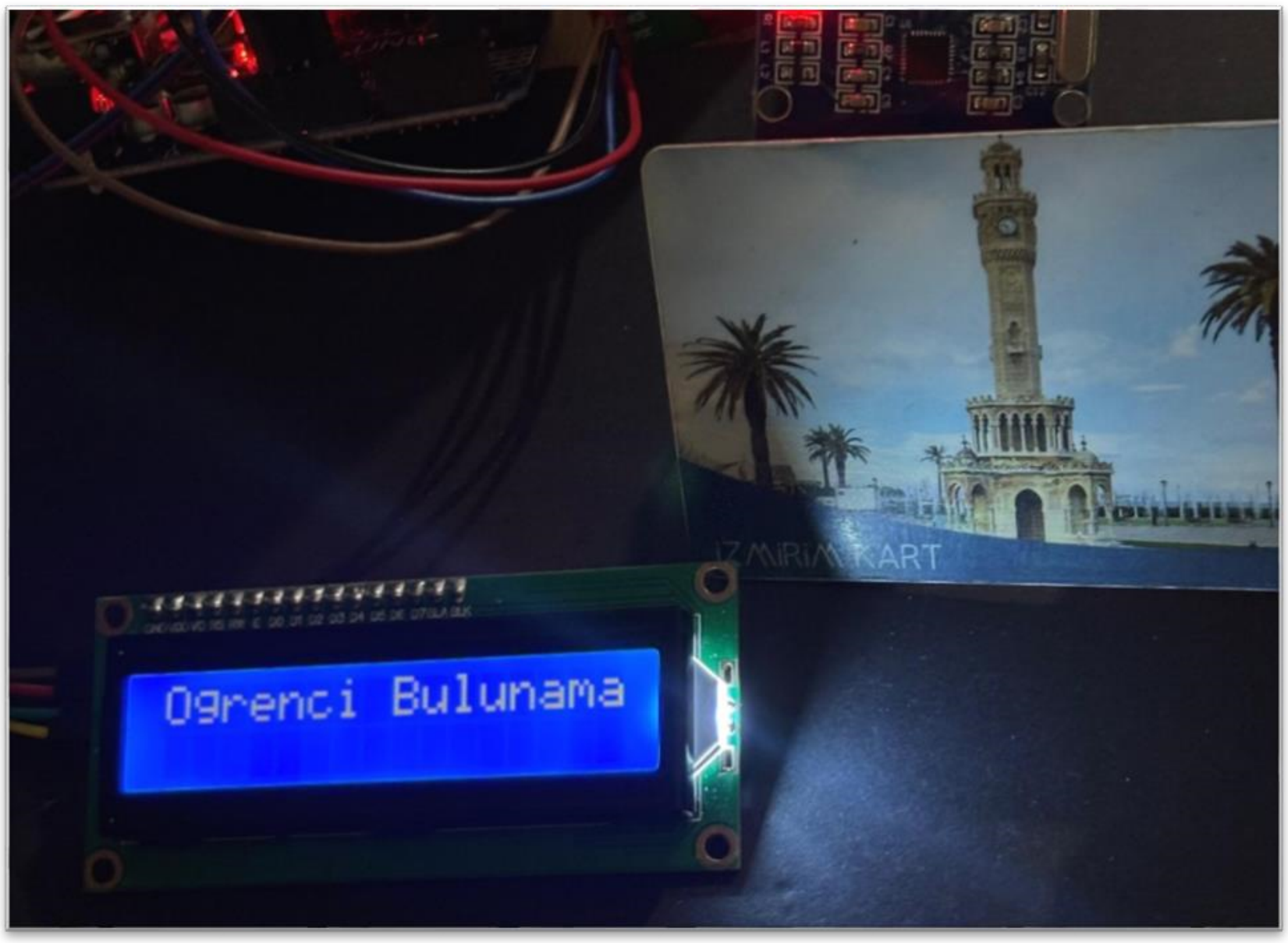

*Şekil 16. Kayıtsız kart okutulması durumunda LCD görüntüsü: "Öğrenci Bulunamadı".*

### *Ağırlık Sensörü Modülü*

Yük hücreleri bilinen referans ağırlıklarla kalibre edildikten sonra sistematik olarak test edilmiştir. ADC değerlerinin kilogram birimine dönüştürülmesinde tutarlı sonuçlar elde edilmiştir. Ortam titreşimlerinin ölçümlerde kararsızlığa yol açtığı gözlemlenmiş; yazılım katmanında uygulanan tolerans tabanlı filtreleme algoritmasıyla bu sorun çözülmüştür.

### *Bluetooth İletişim Modülü*

HC-06 modülü, Arduino ile Python GUI arasındaki veri akışı açısından test edilmiştir. Veri iletiminin genel olarak güvenilir biçimde gerçekleştiği doğrulanmış; zaman zaman gecikme yaşandığı tespit edilmiştir.

Veri gönderim sıklığının optimize edilmesiyle gecikmeler kabul edilebilir düzeye indirilmiştir. Ayrıca Arduino IDE Serial Port monitörü açıkken Python GUI'nin COM portuna erişemediği saptanmış; testlerin yalnızca GUI çalışırken, Serial Port monitörü kapalı durumdayken yürütülmesi gerektiği sonucuna varılmıştır.

***Python GUI Modülü***

Arayüzün tüm işlevsel bileşenleri ayrı ayrı test edilmiştir. Zaman/ders eşleşmesi denetimi, derse kayıtsız kart okutma ve ders saati dışı girişimler dahil olmak üzere hata senaryoları sistematik biçimde sınanmıştır.

Ders saati dışı gerçekleştirilen okutma girişimlerinde Şekil 17'de görülen yanlış saat uyarısı başarıyla tetiklenmiştir. Tüm senaryolarda sistem beklenen davranışı sergilemiştir.

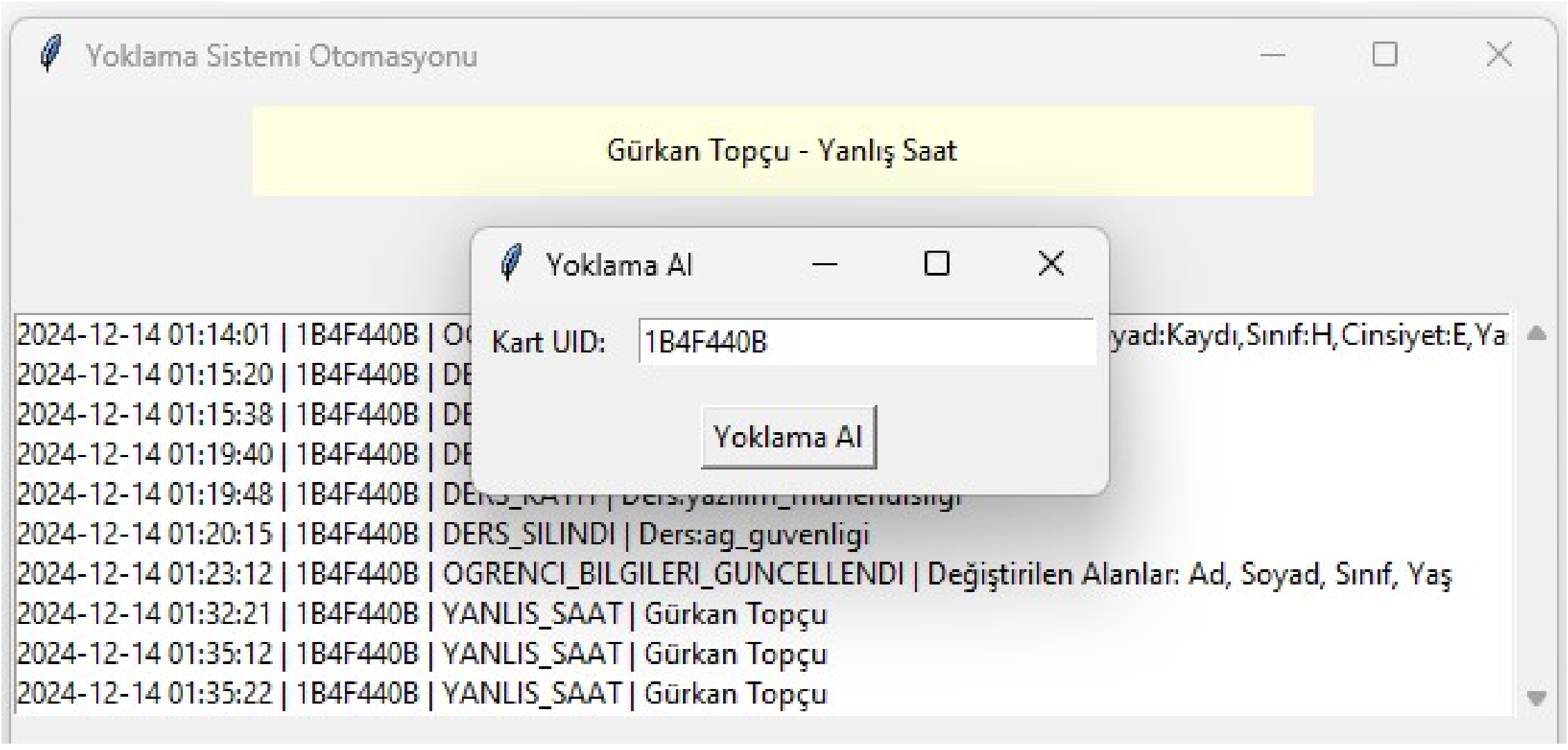

*Şekil 17. Ders saati dışı okutmada "Yanlış Saat" uyarısı.*

## 6.3. Bütünleşik Sistem Testi

Modül testlerinin tamamlanmasının ardından sistem gerçek sınıf düzenine yakın koşullarda bir bütün olarak test edilmiştir. Bütünleşik testler dört temel işlevin sorunsuz biçimde yerine getirildiğini göstermiştir: geçerli kart okutmalarının başarıyla kaydedilmesi, yetkisiz kart girişimlerinin reddedilmesi, ders saati dışı okutmalarda uyarı verilmesi ve ağırlık ile RFID doğrulamasının çapraz kontrolünün tutarlı sonuçlar üretmesi. Şekil 18'de sistemin gerçek çalışması sırasında GUI üzerinde elde edilen yoklama kayıtları görülmektedir.

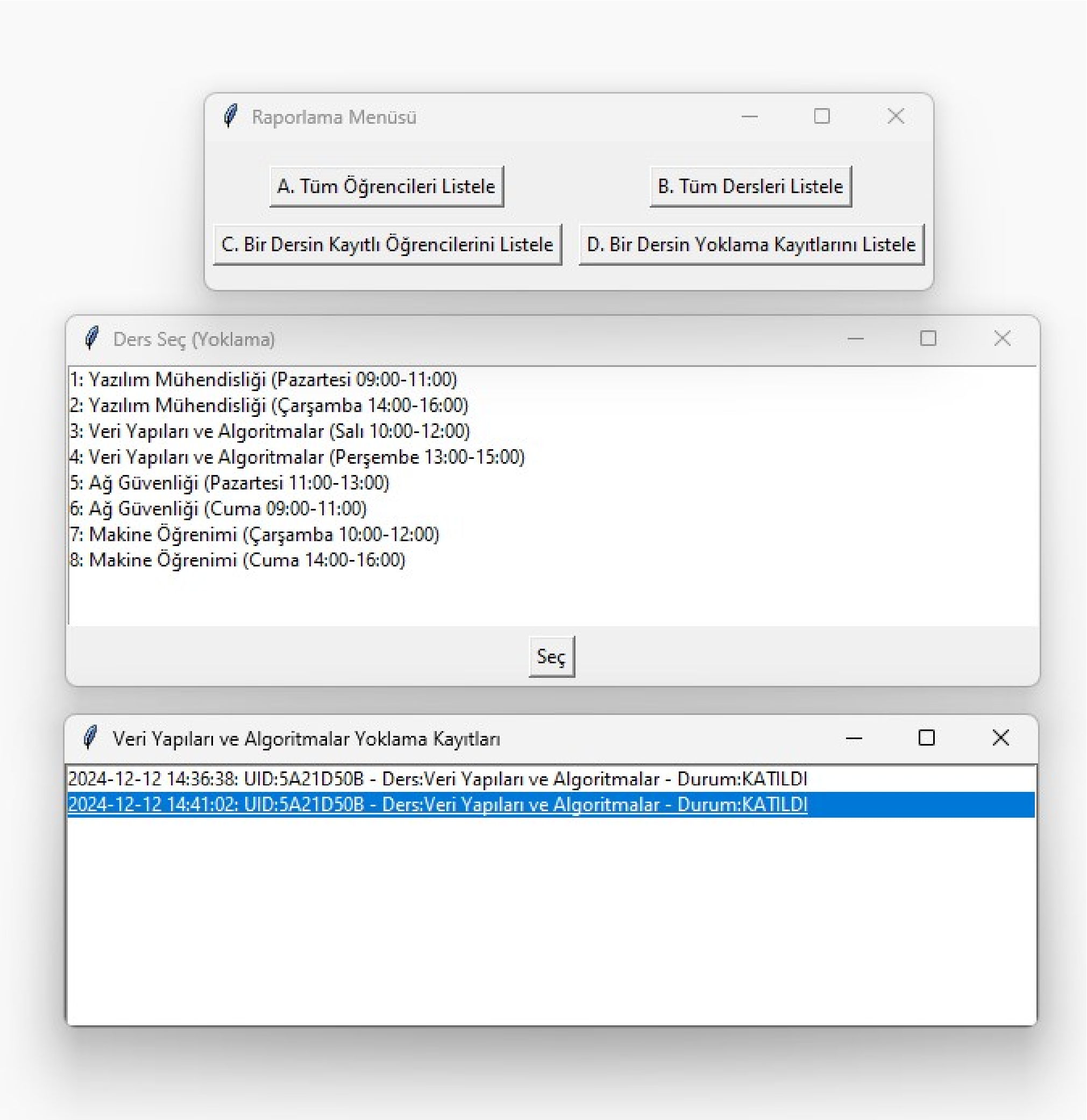

*Şekil 18. Gerçek çalışma sırasında GUI üzerinde elde edilen yoklama kayıtları.*

Sistemin bellek kullanımı ile dosya okuma/yazma işlevleri test süresince kararlı kalmıştır.

## 6.4. Karşılaşılan Sınırlılıklar

Prototip, eş zamanlı yalnızca tek bir sandalyeyi desteklemektedir; çok sandalyeli ortama genişletme ek donanım ve yazılım gerektirmektedir. Ağırlık doğrulama mekanizması, Bölüm 4'teki istatistiksel aralığın dışında kalan bireyler için yanlış uyarı üretme olasılığı barındırmaktadır. Sistem performansı, gerçek bir üniversite sınıfında karşılaşılabilecek titreşim, ısı değişimi ve elektromanyetik parazit koşullarında kapsamlı biçimde test edilmemiştir.

# 7. TARTIŞMA

## 7.1. Önerilen Sistemin Mevcut Yaklaşımlarla Karşılaştırılması

Bu çalışmada geliştirilen sistem, elektronik yoklama literatüründeki biyometrik veri zorunluluğu ve vekâleten yoklama açığı gibi iki köklü sorunu eş zamanlı olarak ele alan bir mimari önermektedir.

Biyometrik tabanlı sistemler yüksek doğruluk oranları raporlamaktadır. Soewito vd. (2016) %95, Dey vd. (2014) ise %94,2 doğruluk oranı bildirmiştir. Bununla birlikte, bu sistemlerin tamamı bireyin kalıcı fiziksel özelliklerinin merkezi olarak işlenmesine veya depolanmasına dayanmaktadır (Kindt, 2013). Önerilen sistem ise herhangi bir biyometrik veri depolamamakta; bu yönüyle GDPR, KVKK ve FERPA bağlamındaki veri koruma yükümlülüklerini önemli ölçüde azaltmaktadır.

Yalnızca RFID kullanan sistemler kart devredildiğinde vekâleten yoklama sorununa karşı yapısal bir güvence mekanizması içermemektedir (Duroc, 2022). Önerilen sistem bu açığı ağırlık sensörü entegrasyonuyla kapatmaktadır. RFID ile biyometrik doğrulamayı birleştiren hibrit yaklaşımlar vekâleten yoklama sorununu büyük ölçüde çözse de biyometrik veri depolama yükümlülüğünü ortadan kaldırmamaktadır. Maliyet açısından, pasif RFID etiketleri ve yük hücrelerinin toplam maliyeti parmak izi okuyucular ve iris tarayıcıların kurulum maliyetinin önemli ölçüde altında kalmaktadır (Want, 2006).

## 7.2. Sınırlılıkların Değerlendirilmesi

Bu çalışmada tanımlanan kilo havuzu mekanizması, çok sandalyeli tam sınıf mimarisi için öngörülen doğrulama mantığını temsil etmektedir. Bununla birlikte, bu makalede sunulan prototip kavram kanıtı düzeyinde tek sandalye ile sınırlandırılmıştır. Bu nedenle, sınıf geneli toplam ağırlık karşılaştırması henüz tam ölçekli donanım üzerinde uygulanmamıştır.

Kaggle kaynaklı 350 bireylik örneklemin farklı coğrafi ve demografik grupları ne ölçüde temsil ettiği belirsizliğini korumaktadır. Örneklemin bir bölümünde ağırlık değerleri, sabit boy varsayımı altında BMI üzerinden türetilmiştir. Bu durum, referans dağılımının kısmen sentetik bir nitelik taşıdığını göstermektedir. Ayrıca, aşırı düşük ya da yüksek kilolu öğrenciler için yanlış uyarı üretilmesi olasılığı, dikkatle değerlendirilmesi gereken bir kullanılabilirlik sorunu oluşturmaktadır. Bunun yanında, prototip aşamasında öğrenci yaş bilgisi simüle edilmiştir. Gerçek kurumsal uygulamada bu verinin üniversite bilgi sistemlerinden otomatik olarak alınması gerekmektedir. Son olarak, sistemin uzun süreli saha performansı henüz kapsamlı ve nicel biçimde değerlendirilmemiştir.

## 7.3. Gelecek Çalışmalar

Çok sandalyeli bir ortamda sistemin ölçeklendirilmesi, gerçek öğrenci yaş verisiyle kurumsal kayıt sistemi entegrasyonunun sağlanması ve nicel bir validasyon çerçevesinin oluşturulması, gelecek araştırmaların öncelikli gündem maddelerini oluşturmaktadır. Mobil uygulama katmanının eklenmesi öğrencilerin devam bilgilerine gerçek zamanlı erişimini mümkün kılacaktır. Makine öğrenimi tabanlı devamsızlık tahmin modellerinin geliştirilmesi, sistemin karar destek süreçlerine daha derin biçimde entegre edilmesine zemin hazırlayabilir.

# 8. SONUÇ

Bu çalışmada, elektronik yoklama alanındaki biyometrik veri zorunluluğu ve RFID tabanlı sistemlerdeki fiziksel varlık doğrulama açığı gibi iki köklü problemi eş zamanlı olarak gideren iki katmanlı bir IoT mimarisi geliştirilmiş ve prototip düzeyinde test edilmiştir. Sistemin özgün katkısı, kimlik doğrulama (RFID) ile fiziksel varlık doğrulama (ağırlık sensörü) görevlerini birbirinden bağımsız iki katmana ayırmasında yatmaktadır. Bu mimari tercih sayesinde kart devredildiğinde gerçekleşen vekâleten yoklama, biyometrik veri depolamak zorunda kalmadan istatistiksel bir referans çerçevesiyle tespit edilebilmektedir.

Çalışmanın dört temel katkısı şöyle özetlenebilir. Birincisi, herhangi bir biyometrik veri depolamayan gizlilik odaklı bir yoklama mimarisi ortaya konulmuştur; anlık ağırlık ölçümü kaydedilmeyip yalnızca bir kerelik karşılaştırmaya tabi tutulmakta, bu yapı GDPR, KVKK ve FERPA bağlamındaki veri koruma yükümlülüklerini yapısal düzeyde azaltmaktadır. İkincisi, RFID kimlik doğrulamasını ağırlık sensörü tabanlı fiziksel varlık denetimiyle bütünleştiren ve kilo havuzu mekanizması üzerine kurulu özgün bir vekâleten yoklama tespit mekanizması tasarlanmıştır. Üçüncüsü, üç Kaggle veri setinden derlenen 350 bireylik örneklem üzerinde yaş ve cinsiyete göre ağırlık dağılımı sistematik biçimde analiz edilmiş; bu analiz hem bireysel sandalye eşik mantığına hem de sınıf geneli kilo havuzu hesaplamasına altlık oluşturmuştur (Kaggle, 2018; Kaggle, 2023; Kaggle, 2024). Dördüncüsü, prototip gerçek sınıf koşullarına yakın bir ortamda test edilmiş; RFID okuma, ağırlık doğrulama, Bluetooth iletişimi ve GUI modüllerinin bütünleşik olarak beklenen işlevleri sergilediği sekiz farklı nitel test senaryosunda gözlemlenmiştir (Tablo 1).

Çalışmanın temel sınırlılıkları şunlardır: prototip tek sandalye ölçeğinde gerçekleştirilmiştir, yaş bilgisi kavram kanıtı aşamasında simüle edilmiştir ve sistematik nicel validasyon metrikleri henüz oluşturulmamıştır. Bununla birlikte, bu çalışma literatürde mevcut sistemlerin kapatamadığı bir tasarım boşluğunu somut bir prototiple ele almış; biyometrik kimlik doğrulamaya bağımlı kalmadan güvenilir bir sınıf yoklama sisteminin hayata geçirilebileceğini ortaya koymuştur. Çok sandalyeli ölçeklendirme, gerçek yaş verisiyle kurumsal entegrasyon ve nicel validasyon çerçevesinin kurulması, bu çalışmanın doğal devamı niteliğindeki gelecek araştırmaların öncelikli hedefleridir.

# KAYNAKLAR